\documentclass[%
 reprint,
superscriptaddress,
 amsmath,amssymb,
 aps,
 pra,
]{revtex4-2}

\usepackage{graphicx}
\usepackage{dcolumn}
\usepackage{bm}
\usepackage{braket}
\usepackage[english]{babel}
\usepackage[nopatch=footnote]{microtype}
\usepackage{xurl}
\usepackage{ragged2e}
\usepackage{xcolor}
\usepackage{soul}
\usepackage[colorinlistoftodos]{todonotes}
\usepackage{titlesec}
\titleformat{\paragraph}
  {\centering\normalfont\small}
  {}{0pt}{}

\titlespacing*{\paragraph}
  {0pt}{1.5ex plus .5ex minus .2ex}{1ex}

\begin{document}
\raggedbottom
\allowdisplaybreaks[1]

\preprint{APS/123-QED}

\title{Connes Distance in Discrete-Time Quantum Walks}

\author{Rayan Trabelsi}
\affiliation{%
Université Paris-Saclay, Inria, CNRS, LMF, 91190 Gif-sur-Yvette, France
}%

\author{A.F. Reyes-Lega}
\affiliation{
Department of Physics, Universidad de los Andes, Cra 1 \#18A-12, 111711, Bogot\'a, Colombia
}%
\author{Pablo Arrighi}
\affiliation{%
Université Paris-Saclay, Inria, CNRS, LMF, 91190 Gif-sur-Yvette, France
}%

\date{\today}

\begin{abstract}
We study the geometric spread of discrete-time quantum walks using Connes' distance formula. 
We introduce its discrete analogue by adapting the continuous spectral metric to the lattice, with the Dirac operator replaced by the unitary step operator.
Focusing first on a $2+1$-dimensional quantum walk, we evaluate the dynamical spread of localized wavepackets to compare propagation in free space against that under a uniform magnetic field.
We find that while the free walk shows linear ballistic growth, gauge-induced phases in the magnetic walk constrain the admissible test operators, suppressing the long-time geometric expansion. Furthermore, we apply this discrete framework to a $1+1$-dimensional plastic quantum walk simulating an inhomogeneous spatial metric. The results demonstrate that the discrete Connes distance naturally captures the underlying spatially varying metric, providing a robust measure of the wavepacket's spread that adapts directly to the local simulated geometry.
\end{abstract}

\maketitle

\section{Introduction}
Discrete-time quantum walks (DTQWs) \cite{Arnault_2021, PhysRevA.97.062111, Arrighi_Nesme_Forets_2014} that simulate the Dirac equation provide a framework for studying relativistic quantum mechanics on a lattice. Discretizing the continuous time evolution of a Dirac particle gives a local unitary operator that defines the dynamics of a quantum walker and offers a discrete setting in which relativistic propagation and geometric notions of distance can be studied simultaneously. 

In the absence of external fields, the spatial structure underlying a quantum walk is described by a commutative position algebra. The introduction of a magnetic field changes the translation structure through the gauge coupling: spatial translations accumulate path-dependent phases, and magnetic translation operators become noncommutative \cite{Zak1964MAGNETICTG}. This provides a natural setting to study how gauge interactions affect the spectral geometry associated with the quantum walk dynamics. 

Noncommutative geometry (NCG) \cite{Connes_1994, Connes1995, Suijlekom_2023a} provides a framework for defining geometric quantities such as distance directly from an algebra and a Dirac operator, without requiring classical points as fundamental objects. In this framework, the geometry is determined by a constraint expressed by the commutator of the observables with the Dirac operator. Because the discrete step operator of the quantum walk converges to this Dirac Hamiltonian in the continuous-time limit, the Connes distance becomes a natural tool for studying the geometric spread of quantum states.

In this work, we investigate the spread of discrete-time Dirac quantum walks, focusing on two physical scenarios: gauge coupling in two spatial dimensions ($2+1$D) and an inhomogeneous spatial metric in one spatial dimension ($1+1$D). 
First, we construct a $2+1$-dimensional quantum walk to compare propagation in free space against propagation under a uniform magnetic field. Analyzing the associated continuous spectral distance, we show that the choice of the algebra plays a crucial role. When using a commutative position algebra, the magnetic potential coupled to the Dirac operator acts as the free Dirac operator. However, for a restricted family of Weyl operators involving both momentum and spatial translations, the magnetic field couples the momentum and translation parameters, which develops a nontrivial kernel.

We next formulate the analogous construction on the lattice, replacing the continuous Dirac operator with the unitary step operator of the quantum walk. Using this discrete framework, we evaluate the dynamical spread of Gaussian wavepackets in the $2+1$-dimensional walk. We find that the distance grows ballistically in the free case, whereas the gauge-induced phases modify the propagation and suppress the long-time growth in the magnetic case.

Finally, to demonstrate how the discrete Connes distance captures more complex physical features, we apply our framework to a $1+1$-dimensional plastic quantum walk \cite{dimolfetta2019quantumwalkcontinuoustimecontinuousspacetime}. This provides a natural setting to study wavepacket dynamics in simulated inhomogeneous spaces, showing that the discrete Connes distance acts as a geometry-sensitive measure that dynamically adapts to spatially varying light speeds.

The paper is organized as follows. Section \ref{sec:dtqw} constructs the discrete Dirac quantum walk in free space and in a uniform magnetic field. Section \ref{ncg} introduces the relevant notions from noncommutative geometry and specifies the observable algebras used throughout the work. Section \ref{continuous} evaluates the continuous spectral distance, specifically comparing the effects of the free and magnetic Dirac operators on the commutative position algebra versus the restricted Weyl family. Section \ref{discrete} develops the analogue discrete formulation. Section \ref{dynamical} applies this construction to the dynamical spectral spread of the free and magnetic $2+1$-dimensional walks and to the $1+1$-dimensional plastic quantum walk. The appendices provide the analytical derivations and numerical implementation details.

\section{\label{sec:dtqw}The Discrete Quantum Walk}\label{quantum-walk}
The evolution generated by the Dirac Hamiltonian $D_{sys}$ is described by $U(t) = e^{-i t D_{sys}}$. A discrete-time quantum walk is obtained by introducing a time step $\epsilon$, such that $t = N\epsilon$, and defining the single-step evolution operator $U_\epsilon = e^{-i \epsilon D_{sys}}$.
\subsection{\label{sec:freedtqw}The Free Particle Evolution}
The Hamiltonian of a free Dirac particle in 2+1D is given by:
\begin{equation*}
    D_{free} = m \sigma_y + \sigma_x p_y + \sigma_z p_x.
\end{equation*}
The single-step evolution operator is: 
\begin{equation*}
    U_\epsilon 
    = e^{-i\epsilon m\sigma_y - \epsilon \sigma_x\partial_y - \epsilon \sigma_z \partial_x }.
\end{equation*}
We define respectively the mass-dependent coin rotation and the conditional spatial shift operators along the $x$-axis and $y$-axis
\begin{equation*}
    M_\epsilon = e^{-im\epsilon \sigma_y}, \quad
    T_x = e^{-\epsilon \sigma_z\partial_x}, \quad
    T_y = e^{-\epsilon \sigma_z\partial_y}.
\end{equation*}
Since the Pauli matrices do not commute, we apply a first order trotterization. Using $\sigma_x = H\sigma_zH$ and setting $W_\epsilon = M_\epsilon  (H T_y H) T_x $, we obtain:
\begin{align}
    U_\epsilon &= M_\epsilon  (H T_y H) T_x +\mathcal{O}(\epsilon^2) \nonumber \\
               &= W_\epsilon +\mathcal{O}(\epsilon^2).
\end{align}
These shift operators act on the joint Hilbert space $\mathcal{H} = \mathcal{H}_p \otimes \mathcal{H}_c$ of the particle's position and its internal coin state. 
In order to discretize space, we couple the spatial lattice spacing to the time step such that $\Delta x = \Delta y = \epsilon$
and we map the physical coordinates $(x, y)$ to integer indices $(n, m)$ such that $x = n\epsilon$ and $y = m\epsilon$. The basis states are denoted $\ket{n, m} \otimes \ket{c}$.
The action of $T_x$ and $T_y$ translates the particle conditioned on its coin state:
\begin{align*}
    T_x \ket{n, m} \otimes \ket{0} &= e^{- \epsilon \partial_x} \ket{n, m} \otimes \ket{0} = \ket{n+1, m} \otimes \ket{0}, \\
    T_x \ket{n, m} \otimes \ket{1} &= e^{+ \epsilon \partial_x} \ket{n, m}\otimes \ket{1} = \ket{n-1, m} \otimes \ket{1}, \\
    T_y \ket{n, m} \otimes \ket{0} &= e^{-\epsilon \partial_y} \ket{n, m} \otimes \ket{0} = \ket{n, m+1} \otimes \ket{0}, \\
    T_y \ket{n, m} \otimes \ket{1} &= e^{+\epsilon \partial_y} \ket{n, m} \otimes \ket{1}= \ket{n, m-1} \otimes \ket{1}.
\end{align*}
Since $[p_x,p_y] = 0$, the conditional shifts commute ($[T_x,T_y]=0$). Thus, the free quantum walk is generated by commuting translations. As we show in the next section, coupling the walk to a magnetic field modifies this translation algebra to a noncommutative one.

\subsection{\label{magdtqw}Gauge-Coupled Evolution in a Magnetic Field}
We now introduce a magnetic field perpendicular to the plane, coupled to the Dirac particle of charge $1$ through a gauge vector potential $\vec{A} = (A_x(y), A_y(x))$ where $A_x(y)$ and $A_y(x)$ are linear functions of their respective arguments. The magnetic field is uniform and is given by: 
\begin{equation*}
    B=\partial_xA_y-\partial_yA_x.
\end{equation*}
The momentum operators are modified to $\Pi_j = p_j - A_j$, giving the Hamiltonian:
\begin{equation*}
    D_B = m \sigma_y + \sigma_x \Pi_y + \sigma_z \Pi_x,
\end{equation*}
where
\begin{align*}
    \Pi_x &= p_x - A_x(y), \\
    \Pi_y &= p_y - A_y(x).
\end{align*}
They satisfy
\begin{equation*}
    [\Pi_x,\Pi_y]=iB.
\end{equation*}
In the limit where $A_x(y) = A_y(x) = 0$, we recover the free particle Hamiltonian. 

The single-step evolution operator becomes:
\begin{align}
    U_\epsilon 
    &= e^{-i\epsilon(m \sigma_y + \sigma_x \Pi_y + \sigma_z \Pi_x)} \nonumber \\
    &= e^{-im\epsilon \sigma_y} e^{-i\epsilon \sigma_x \Pi_y} e^{-i \epsilon \sigma_z \Pi_x} + \mathcal{O}(\epsilon^2) \nonumber \\
    &= M_\epsilon (H T_y^B H) T_x^B + \mathcal{O}(\epsilon^2) \nonumber \\
    &= W_\epsilon^B + \mathcal{O}(\epsilon^2) .
\end{align}
The mass operator $M_\epsilon$ remains identical to the free particle case, because the magnetic field couples only to the momentum and does not affect the mass term. However, the spatial shift operators now depend on the gauge potential:
\begin{align*}
    T_x^B &= e^{-i \epsilon \sigma_z \Pi_x} = e^{-\epsilon \sigma_z \partial_x + i\epsilon A_x(y) \sigma_z}, \\
    T_y^B &= e^{-i \epsilon \sigma_z \Pi_y} = e^{-\epsilon \sigma_z \partial_y + i\epsilon A_y(x) \sigma_z}.
\end{align*}
Their action on localized states accumulates position-dependent phases:
\begin{align*}
    T_x^B \ket{n, m} \otimes \ket{0} &= e^{i\epsilon A_x(m\epsilon)} \ket{n+1, m} \otimes \ket{0}, \\
    T_x^B \ket{n, m} \otimes \ket{1} &= e^{-i\epsilon A_x(m\epsilon)} \ket{n-1, m} \otimes \ket{1}, \\
    T_y^B \ket{n, m} \otimes \ket{0} &= e^{i\epsilon A_y(n\epsilon)} \ket{n, m+1} \otimes \ket{0}, \\
    T_y^B \ket{n, m} \otimes \ket{1} &= e^{-i\epsilon A_y(n\epsilon)} \ket{n, m-1} \otimes \ket{1}.    
\end{align*}
Unlike the free particle case, the modified shift operators no longer commute ($[T_x^B, T_y^B] \neq 0$) because $[\Pi_x, \Pi_y] \neq 0$ ($[p_x, A_y(x)] \neq 0$ and $[p_y, A_x(y)] \neq 0$). When applied in sequence on a basis state $\ket{n, m} \otimes \ket{0}$:
\begin{align*}
    &T_y^B T_x^B \ket{n, m} \otimes \ket{0} \\
    &\quad = e^{i\epsilon [A_x(m\epsilon) + A_y((n+1)\epsilon)]}
      \ket{n+1, m+1} \otimes \ket{0}, \\
    &T_x^B T_y^B \ket{n, m} \otimes \ket{0} \\
    &\quad = e^{i\epsilon [A_y(n\epsilon) + A_x((m+1)\epsilon)]}
      \ket{n+1, m+1} \otimes \ket{0}.
\end{align*}
It follows that 
\begin{equation*}
    T_y^B T_x^B = e^{iB\epsilon^2} T_x^B T_y^B.
\end{equation*}
In the presence of a non-zero magnetic field and when $B\epsilon^2$ is not a multiple of $2\pi$, $T_y^B T_x^B \neq T_x^B T_y^B$. Thus, the gauge coupling forces the translation algebra to become noncommutative \cite{Zak1964MAGNETICTG}. In the following section,
we investigate how this algebraic structure enters the construction of
a spectral metric.

\section{Spectral Geometry and Connes Distance} \label{ncg}
\subsection{Spectral Triples and Connes (Spectral) Distance}
Classical geometry is built on the concept of a space formed by a set of points. The Gelfand-Naimark theorem~\cite{varilly2006introductionnoncommutativegeometry} shows that a  topological space (Hausdorff, locally compact) $X$ can be 
reconstructed from the commutative algebra $C(X)$ of continuous functions defined on it. This algebra naturally carries the structure of a (commutative) $C^*$-algebra. Moreover, topological properties and constructions on $X$ translate into purely algebraic statements about $C(X)$. This has led to the idea of considering general noncommutative $C^*$-algebras as ``noncommutative topological spaces'' (for a physically motivated introduction to $C^*$-algebras, see~\cite{Reyes-Lega2016}).

Noncommutative geometry provides a vast generalization of this idea, where notions such as distance, curvature, or topological invariants keep their meaning, even without an underlying manifold of points.
The transition from classical to noncommutative geometry is, in a certain sense, similar to the transition from classical mechanics to quantum mechanics, for which observables generally do not commute. 
Thus, if we replace the commutative algebra of functions on a space with a noncommutative one, we lose the classical notion of sharply localized points, in a way reminiscent of the uncertainty principle.

In Connes' approach to noncommutative geometry, the properties of a space are encoded by a so-called spectral triple $(\mathcal A,\mathcal H,D)$. $\mathcal A$ is a dense $*$-subalgebra of a $C^*$-algebra $A$ and $\mathcal H$ is a 
Hilbert space carrying a representation
\begin{equation*}
    \pi: A\longrightarrow\mathcal B(\mathcal H),
\end{equation*}
of $A$ by bounded operators. 
$D$ is a self-adjoint operator on $\mathcal H$, subject to the condition
\begin{equation*}
    [D,\pi(a)]\in\mathcal B(\mathcal H),
    \qquad a\in\mathcal A.
\end{equation*}
For a unital spectral triple, the operator  $(1+D^2)^{-1/2}$ is required to be compact. Additional structures can also be included in the definition: a $\mathbb Z_2$-grading operator, relevant for even-dimensional geometries, and  a real structure (charge-conjugation operator) needed to describe spin (as opposed to $\mbox{spin}^c$) structures in either parity. Neither  will  be needed in the present work. 

In the classical, commutative case,  $\mathcal A$ is the  algebra of smooth functions on a \emph{spin  (or spin$^c$) manifold} $M$, $\mathcal A =C^\infty(M)$, $A=C(M)$ and $D$ is a globally defined Dirac operator (or its $\mbox{spin}^c$ counterpart)~\cite{hijazi,GraciaBondia2001}.

As a consequence of Connes' Spin Manifold Theorem, the metric structure of $M$ can be recovered from the Dirac operator \cite{varilly2006introductionnoncommutativegeometry,GraciaBondia2001}.

In the noncommutative case, the algebra $A$ provides the observables used to describe the space, while the Dirac operator $D$ 
is now \emph{used} in order to define a metric structure.
In this context,
the classical notion of points is lost, but it can be replaced by that  of algebraic states: positive, normalized linear functionals
\begin{equation*} 
    \varphi: A\longrightarrow\mathbb C, \qquad \varphi(a^\ast a)\geq0, \qquad \varphi(\mathbb I)=1. 
\end{equation*}
Instead of evaluating a classical function at a specific point, we evaluate an operator $a \in A$ through its expectation value in a specific state. This matches the physical interpretation of quantum systems, where states describe distributions rather than exact positions.

To measure the distinguishability between two states, $\varphi$ and $\psi$, we use the Connes distance~\cite{Connes_1994}:
\begin{equation}\label{eq:connes-distance}
    d_D(\varphi,\psi) = \sup_{a\in\mathcal{A}} \left\{ |\varphi(a)-\psi(a)| : \|[D,\pi(a)]\| \leq 1 \right\}.
\end{equation}
Since the Dirac operator acts as a generalized first order differential operator, the commutator norm imposes a ``speed limit'' on the space, it bounds how fast a test observable can vary. Therefore, the constraint can be seen as a Lipschitz continuity bound. 

To build physical intuition, we recall a standard commutative NCG example \cite{Suijlekom_2023a, Connes1995}. On a continuous one-dimensional manifold where $D = -i\partial_x$, the commutator $[D, \pi(f)]$ gives the spatial gradient, which restricts test functions to a maximum slope of unity and recovers the standard geodesic distance.

As defined above, the Connes distance depends not only on the algebra, but also on the Dirac operator and the representation $\pi$. This fact becomes particularly clear in the finite-dimensional case, as explained in~\cite{Suijlekom_2023a}. It is important to note that a noncommutative algebra alone does not automatically guarantee a non-Euclidean metric. The operator $D$ must ``see'' the noncommutative directions through the commutator $[D, \pi(a)]$~\cite{2020}. In a similar way, for a commutative algebra of scalar functions or operators, the magnetic gauge potential commutes with the algebra and does not modify the commutator's norm. As we show in the following sections, coupling a uniform magnetic field to the Dirac operator and using a noncommutative algebra interlinks position and momentum translations in the Lipschitz constraint which modifies the metric structure of the quantum walk.

\subsection{Choice of the Algebra}
Our main interest in this work lies in the dual role played by the Dirac operator: on the one hand, it is the generator of the (discrete) time evolution for the quantum walk and, on the other hand, it induces a metric structure on the space of quantum states (via the spectral distance formula). Although the spectral distance appears naturally
in the setting of spectral triples, it may be used in a purely quantum mechanical context (as is done here), where one is not necessarily working with a full spectral triple. 

In any case, the distance formula \eqref{eq:connes-distance} is quite sensitive to the interplay between the algebra and the Dirac operator. In order to illustrate this fact, we consider two classes of observables. This will allow us to distinguish the effect of the algebra from that of the Dirac operator.

\subsubsection{Commutative Position Algebra} \label{comm-algebra}
We consider an algebra $\mathcal{A}_{\text{com}}$ of bounded functions of position represented on $\mathcal H = L^2(\mathbb R^2)\otimes\mathbb C^2$ by:
\begin{equation*}
    \pi_{\text{com}}(f) = M_f \, \otimes \mathbb I_2, \qquad f \in \mathcal A_{\text{com}}.
\end{equation*}
This algebra contains only scalar functions of position which naturally commute with the position-dependent gauge potentials, a property that will be used when evaluating the constraint in Connes' distance in Section \ref{comm-dirac}.

\subsubsection{Restricted Weyl Operator Family}
To study observables involving both position and momentum translations, we consider the Weyl operators
\begin{equation*}
    T(k,q) = e^{i(k\cdot \hat{r} + q \cdot \hat{p})}, \qquad k,q \in \mathbb R^2,
\end{equation*}
where
\begin{equation*}
    [\hat{r}_j,\hat{p}_l] = i \delta_{jl}.
\end{equation*}
Using the Baker-Campbell-Hausdorff formula, they can be written as
\begin{equation*}
    T(k,q) =
    e^{ik\cdot\hat{r}}
    e^{iq\cdot\hat{p}}
    e^{\frac{i}{2}k\cdot q}.
\end{equation*}
These operators generate the Weyl algebra, a noncommutative algebra containing both position and momentum translations. In this work, however, we do not evaluate the Connes distance over the full Weyl algebra but over the family of individual Weyl operators. Therefore, the resulting distance is a Weyl-restricted Connes distance, which provides a lower bound on the Connes distance associated with the full Weyl algebra.

\subsection{States Used in the Distance Calculation}
We evaluate the distance between localized Gaussian states. We consider the normalized wavepacket
\begin{equation*}
    \zeta_{r_0}(r) =
    \frac{1}{\sqrt{\pi}\sigma}
    e^{-\frac{| r- r_0|^2}{2\sigma^2}},
\end{equation*}
centered at $r_0\in\mathbb R^2$. Its probability density is
\begin{equation*}
    |\zeta_{r_0}(r)|^2
    =
    \frac{1}{\pi\sigma^2}
    e^{-\frac{|r-r_0|^2}{\sigma^2}},
\end{equation*}
and each cartesian coordinate has variance $\sigma^2/2$.

We fix a normalized coin spinor $\chi\in\mathbb C^2$ and define the
corresponding state vector in
$L^2(\mathbb R^2)\otimes\mathbb C^2$ as
\begin{equation*}
    \ket{\psi_{r_0}} = \ket{\zeta_{r_0}}\otimes\ket{\chi}.
\end{equation*}
This vector defines a state on the observable algebra through
\begin{equation}
    \omega_{r_0}(a) =
    \braket{\psi_{r_0} | \pi(a) | \psi_{r_0}}, \qquad a\in\mathcal A.
\end{equation}

\section{Continuous Connes Distance} \label{continuous}

\subsection{Commutative Algebra with the Magnetic Dirac Operator} \label{comm-dirac}
We first restrict our observables to the commutative algebra of purely spatial functions, $\mathcal{A}_{\text{com}} = \{f(\hat{r})\}$ and we evaluate the commutator's norm using the magnetic Dirac operator $D_B$.
Since the gauge potentials $A_j(\hat{r})$ are only functions of position, they commute with the test observable $f(\hat{r})$. Therefore, the magnetic field contributions disappear from the commutator:
\begin{align*}
    [D_B,\pi_{\text{com}}(f)] &= \sigma_z [p_x - A_x, f] + \sigma_x [p_y - A_y, f] \\
                &= \sigma_z [p_x, f] + \sigma_x [p_y, f] \\
                &= -i (\sigma_z \partial_x f + \sigma_x \partial_y f).
\end{align*}
The commutator's norm is therefore independent of the magnetic field:
\begin{equation*}
    \| [D_B,\pi_{\text{com}}(f)] \| = \sup_{r \in \mathbb R^2} \sqrt{|\partial_x f(r)|^2 + |\partial_y f(r)|^2}.
\end{equation*}
For two Gaussian states $\omega_{r_1}$ and $\omega_{r_2}$, the constraint $\|\nabla f\| \le 1$ restricts the test functions to those with a maximum gradient of unity. The supremum is approached by bounded $1$-Lipschitz truncations of the linear function $f(\hat{r}) = \frac{r_1 - r_2}{\|r_1 - r_2\|} \cdot \hat{r}$. Taking the limit of these truncations, the Connes distance is the standard Euclidean distance (see Appendix \ref{euclidean1}):
\begin{equation}
    d_{D_B}(\omega_{r_1}, \omega_{r_2}) = \|r_1 - r_2\|_2.
\end{equation}
This confirms that if the algebra is commutative, changing the Dirac operator alone from $D_\text{free}$ to $D_B$ does not change the metric. \\
The recovery of Euclidean distance for localized translated states is consistent with spectral-distance results for coherent states in noncommutative geometries~\cite{MartinettiTomassini2013}.

\subsection{Weyl Operators with the Free Dirac Operator} \label{noncomm-freedirac}
We now expand our test observables to the noncommutative family of Weyl operators, $T(k,q) = e^{i(k\cdot \hat{r} + q\cdot\hat{p})}$, but couple them to the free Dirac operator $D_\text{free}$.

Since the Weyl operators act as the identity on the coin space, the mass term vanishes from the commutator. Using the canonical commutation relation $[p_j, e^{ik_j \hat{r}_j}] = k_j e^{ik_j \hat{r}_j}$, the commutator evaluates to:
\begin{equation*}
    [D_\text{free}, \pi(T)] = (\sigma_z k_x + \sigma_x k_y) T(k,q),
\end{equation*}
and because $T(k,q)$ is unitary, the norm of the commutator simplifies to:
\begin{equation*}
    \|[D_\text{free}, \pi(T)]\| = \sqrt{k_x^2 + k_y^2} = \|k\|_2.
\end{equation*}
It is independent of the spatial translation parameter $q$. The free Dirac operator does not detect displacements generated by $q$, making the space degenerate in that algebraic direction.

The Weyl expectation is given by
\begin{equation*}
    \omega_{r_1}(T) = e^{ik\cdot r_1} e^{-\frac{\sigma^2 \|k\|^2}{4}-\frac{\|q\|^2}{4\sigma^2}}.
\end{equation*}
To evaluate the restricted Connes distance, the difference in expectation values is 
\begin{equation*}
    |\omega_{r_1}(T) - \omega_{r_2}(T)| = e^{-\frac{\sigma^2 \|k\|^2}{4}-\frac{\|q\|^2}{4\sigma^2}} |e^{ik\cdot r_1} - e^{ik\cdot r_2}|.
\end{equation*}

For fixed $k$, the optimum selects $q=0$, as this maximizes the Gaussian overlap. Therefore, the distance reduces to the supremum over simple plane waves ($e^{ik\cdot \hat{r}}$), giving the Euclidean lower bound:
\begin{equation*}
    d_{D_\text{free}}(\omega_{r_1}, \omega_{r_2}) \ge \|r_1 - r_2\|.
\end{equation*}
To get the equality, we take the limit when $k \to 0$ and obtain (see Appendix \ref{weyl_free_proof})
\begin{equation}
    d_{D_\text{free}}(\omega_{r_1}, \omega_{r_2}) = \|r_1 - r_2\|_2.
\end{equation}

This result reveals a geometric limitation of the free Dirac operator. Because $D_\text{free}$ depends only on the mass term and the momentum operators, it commutes with the momentum translations generated by $q$. The metric constraint therefore imposes no constraint on $q$. 
However, because we evaluate the distance between purely spatially separated wavepackets, the supremum naturally selects $q=0$ to prevent the exponential suppression of the state overlap. This optimal choice projects the Weyl family back onto its commutative spatial subfamily ($e^{ik\cdot \hat{r}}$).

As we demonstrate in the following section, introducing a gauge potential breaks this degeneracy by interlinking position and momentum.

\subsection{Weyl Operators with the Magnetic Dirac Operator} \label{noncomm-dirac}
We evaluate the gauge-coupled Dirac operator $D_B$ against the Weyl operators $T(k,q)$.
The commutator now includes the position-dependent gauge potentials:
\begin{equation*}
    [\Pi_j, T(k,q)] = [p_j, T] - [A_j(\hat{r}), T].
\end{equation*}
The translation generator $e^{iq\cdot \hat{p}}$ shifts the spatial argument of the gauge potential:
\begin{equation*}
    T(k,q) A_j(\hat{r}) T(k,q)^\dagger = A_j(\hat{r} + q).
\end{equation*}
This generates an additional term in the commutator:
\begin{equation*}
    [\Pi_j, T(k,q)] = \left( k_j - A_j(\hat{r}) + A_j(\hat{r} + q) \right) T(k,q).
\end{equation*}
Choosing the Landau gauge $\vec{A} = (-By, 0)$ to represent a uniform magnetic field $B$ perpendicular to the plane, the shift differences become exact linear terms:
\begin{align*}
    -A_x(\hat{y}) + A_x(\hat{y} + q_y) &= -Bq_y, \\
    -A_y(\hat{x}) + A_y(\hat{x} + q_x) &= 0.
\end{align*}
The spatial coordinates cancel out, making the commutator bounded. The commutator's norm becomes (see Appendix \ref{weyl-magnetic-dirac-proof})
\begin{equation*}
    \|[D_B, \pi(T)]\| = \sqrt{(k_x - Bq_y)^2 + k_y^2}.
\end{equation*}
Unlike the free-particle case, the uniform magnetic field couples the momentum parameter $k_x$ and the spatial translation parameter $q_y$ inside the constraint. The algebra permits operators with spatial translations $q_y$ provided they are compensated by a momentum shift $k_x \approx Bq_y$. At exact compensation, $k_x = Bq_y$ and $k_y=0$, the commutator norm vanishes. 
For these zero-seminorm operators, the expectation difference between two Gaussian states centered at $r_1$ and $r_2$ is proportional to $|e^{i B q_y r_{1x}} - e^{i B q_y r_{2x}}|$.
Whenever the two states have different x-coordinates, there exists an allowed $q_y$ such that the phase difference satisfies $Bq_y(r_{1x}-r_{2x}) \notin 2\pi\mathbb{Z}$, making the expectation difference non-zero. The test operator can then be scaled by an arbitrary factor $c \to \infty$ while preserving the zero-norm constraint, which implies:
\begin{equation}
    r_{1x} \neq r_{2x} \implies d_{D_B}(\omega_{r_1}, \omega_{r_2}) = \infty.
\end{equation}
Infinite Connes distances are known phenomena in noncompact and noncommutative spectral geometries \cite{Cagnache_2011}.
They indicate a breakdown of the spectral metric along specific algebraic directions.

\section{Discrete Connes Distance} \label{discrete}
The Connes distance has also been studied directly on discrete lattices and graphs~\cite{DimakisMuellerHoissen1998,Besnard2021}. However, to apply this metric to the dynamics of the quantum walk, we must map our continuous spectral geometry onto the discrete lattice.

In this discrete framework, the Hilbert space is restricted to $\mathcal{H} = \ell^2(\mathbb{Z}^2) \otimes \mathbb{C}^2$. 
The quantum walk is governed by the unitary step operator $W_\epsilon$ rather than the continuous Dirac Hamiltonian $D$, which requires us to adapt the metric constraint. The walk operator is generated by the Hamiltonian: $W_\epsilon = \mathbb{I} - i\epsilon D + \mathcal{O}(\epsilon^2)$. Therefore, the discrete commutator of an observable with the walk operator is:
\begin{equation} \label{self-adjoint}
    [W_\epsilon, \pi(a)] \approx -i\epsilon [D, \pi(a)].
\end{equation}
Hence, the continuous Lipschitz bound $\|[D_B, \pi(a)]\| \le 1$ translates to the discrete constraint:
\begin{equation} \label{eq:discrete_constraint}
    \frac{1}{\epsilon} \left\| [W_\epsilon, \pi(a)] \right\| \le 1.
\end{equation}
While $W_\epsilon$ is a unitary operator rather than a self-adjoint Dirac operator, Eq. \ref{self-adjoint} shows that the resulting walk-induced seminorm recovers the standard Connes distance in the continuous-time limit.

The discrete Connes distance is evaluated in this framework by taking the supremum of expectation differences between two states $\omega_1$ and $\omega_2$ over all observables $a$ satisfying Eq.~(\ref{eq:discrete_constraint}): 
\begin{equation}
    d_{W_\epsilon}(\omega_1,\omega_2) = \sup_{a\in\mathcal{A}} \left\{ |\omega_1(a)-\omega_2(a)| : \|[W_\epsilon, \pi(a)]\| \leq \epsilon \right\}.
\end{equation}

\subsection{Discrete Constraint Evaluation}
In the following, we map the continuous noncommutative analysis of Sec.\ref{noncomm-dirac} to the discrete lattice. To do so, we evaluate the constraint using the walk operator $W_\epsilon^B$ generated by the magnetic Dirac operator $D_B$. Its algebraic expression is given by:
\begin{equation*}
    W_\epsilon^B = C_2 \cdot D_y \cdot T_y \cdot C_1 \cdot D_x \cdot T_x,
\end{equation*}
where the coin operators are
\begin{equation*}
    C_1 = H, \qquad C_2 = M_\epsilon H.
\end{equation*}
The position-dependent magnetic phases and conditional translations are:
\begin{align*}
    D_x &= \sum_{r} \ket{r}\bra{r} \otimes e^{i \epsilon A_x(r) \sigma_z}, \\
    D_y &= \sum_{r} \ket{r}\bra{r} \otimes e^{i \epsilon A_y(r) \sigma_z}, \\
    T_x &= \sum_{r} \left(\ket{r+e_x}\bra{r} \otimes P_{1} + \ket{r-e_x}\bra{r} \otimes P_{-1} \right), \\ 
    T_y &= \sum_{r} \left(\ket{r+e_y}\bra{r} \otimes P_{1} + \ket{r-e_y}\bra{r} \otimes P_{-1} \right),
\end{align*}
with $r = (x,y) \in (\epsilon\mathbb{Z})^2$ denoting the discrete lattice nodes, $e_x = (\epsilon,0)$ and $e_y=(0,\epsilon)$, and the coin projectors $P_{1} = \ket{0}\bra{0}$ and $P_{-1} = \ket{1}\bra{1}$.

When it acts on a basis state $\ket{r} \otimes \ket{c}$, the walk applies a superposition of jumps $\delta \in \mathcal{N} = \{(\pm \epsilon, \pm \epsilon)\}$, and a coin matrix $U_\delta(r)$:
\begin{equation*}
    W_\epsilon^B \big( \ket{r} \otimes \ket{c}\big) = \sum_{\delta \in \mathcal{N}} \ket{r + \delta} \otimes U_\delta(r) \ket{c},
\end{equation*}
with $\delta = s_x e_x + s_ye_y$ for $s_x, s_y \in \{-1, 1\}$ and $U_\delta(r)$ is given by:
\begin{equation} \label{qw-action-coin}
    U_\delta(r) = M_\epsilon H \, e^{i \epsilon A_y(r + \delta) \sigma_z} P_{s_y} \, H \, e^{i \epsilon A_x(r + s_x e_x) \sigma_z} P_{s_x}.
\end{equation}

We construct the discrete analogue of the Weyl observable family, composed of observables $a$ whose representation on the Hilbert space is:
\begin{equation*}
    \pi(a) = V_k S_q \otimes \mathbb I_2,
\end{equation*}
where $k=(k_x,k_y)$ is the momentum parameter, $q=(q_x,q_y) \in (\epsilon\mathbb{Z})^2$ is a discrete translation vector, and
\begin{align*}
    S_q \ket{r} &= \ket{r + q}, \\
    V_k \ket{r} &= e^{i k \cdot r} \ket{r}.
\end{align*}
In order to evaluate the commutator $[W_\epsilon^B,\pi(a)]$, we study its action on a basis state $\ket{r} \otimes \ket{c}$. 
First
\begin{align*}
    \pi(a) \big( \ket{r} \otimes \ket{c} \big) &= e^{ik \cdot (r+q)} \ket{r + q} \otimes \ket{c}, \\
    W_\epsilon \, \pi(a) \big( \ket{r} \otimes \ket{c} \big)
      &= e^{ik \cdot (r+q)} \sum_{\delta \in \mathcal{N}} \ket{r + q + \delta} \\
      &\qquad \otimes U_\delta(r + q) \ket{c}.
\end{align*}
Then
\begin{align*}
    W_\epsilon \big( \ket{r} \otimes \ket{c} \big) &= \sum_{\delta \in \mathcal{N}} \ket{r + \delta} \otimes U_\delta(r) \ket{c}, \\
    \pi(a)\, W_\epsilon \big( \ket{r} \otimes \ket{c} \big)
      &= \sum_{\delta \in \mathcal{N}} e^{ik \cdot (r + \delta + q)} \\
      &\qquad \times \ket{r + \delta + q} \otimes U_\delta(r) \ket{c}.
\end{align*}
Therefore, subtracting these two results:
\begin{align*}
    [W_\epsilon, \pi(a)] \big( \ket{r} \otimes \ket{c} \big) = \;&e^{ik \cdot (r+q)} \sum_{\delta \in \mathcal{N}} \ket{r + q + \delta} \\ 
    &\otimes \Big( U_\delta(r + q) - e^{ik \cdot \delta} U_\delta(r) \Big) \ket{c}.
\end{align*}
We define
\begin{align*}
    K_\delta(r) &= U_\delta(r+q)-e^{ik\cdot\delta}U_\delta(r), \\
    Z_{k,q} &= \sum_r e^{ik\cdot(r+q)}\ket{r}\bra{r}\otimes\mathbb I_2, \\
    M_{K_\delta} &= \sum_r \ket{r}\bra{r}\otimes K_\delta(r).
\end{align*}
From its action on the basis states, we can reconstruct the commutator as a full operator on the Hilbert space:
\begin{align*}
    [W_\epsilon^B, \pi(a)] 
    &= \sum_{r, \delta} e^{ik \cdot (r+q)} \ket{r + q + \delta}\bra{r} \otimes K_\delta(r) \\
    &= S_q \ \sum_{\delta \in \mathcal{N}} S_\delta \, Z_{k,q} \, M_{K_\delta}.
\end{align*}
Thus, the condition $\|[W_\epsilon, \pi(a)]\| \le \epsilon$ becomes:
\begin{align}
\| [W_\epsilon^B, \pi(a)] \| 
    &= \left\| S_q \left( \sum_{\delta \in \mathcal{N}} S_\delta \, M_{K_\delta} \right) Z_{k,q} \right\| \nonumber \\
    &= \left\| \sum_{\delta \in \mathcal{N}} S_\delta \, M_{K_\delta}  \right\| \\
    &\leq \epsilon. \nonumber
\end{align}

In the absence of the magnetic field, $A_x = A_y = 0$, which implies $U_\delta(r) = M_\epsilon H P_{s_y} H P_{s_x}$. In this case, $U_\delta$ is independent from the position $r$, and because $U_\delta(r+q) = U_\delta(r)$, $K_\delta (r) = U_\delta (r) (1 - e^{ik \cdot \delta} )$.

The translation parameter $q$ disappears from the commutator's norm, reducing the bound to a constraint on $k$.
 
\subsection{Metric Breakdown in the Discrete Magnetic Walk}
In the presence of the magnetic field, $U_\delta(r+q) \neq U_\delta(r)$.
In the continuous framework, the momentum $k_x$ compensates for the magnetic shift $Bq_y$. However, on the discrete lattice, the magnetic phase is coupled to the coin through $\sigma_z$, which splits the gauge shift into two opposite chiralities $+Bq_y$ and $-Bq_y$.

To determine how this discrete chiral structure affects the metric geometry, we evaluate the restricted discrete Connes distance between two localized Gaussian states $\ket{\zeta_{0}}$ and $\ket{\zeta_{1}}$ centered at lattice positions $r_0$ and $r_1$, as an analogue to the continuous analysis in Sec. \ref{noncomm-dirac}. We scale the Weyl operators by an arbitrary positive constant $c > 0$, of the form:
\begin{equation*}
    \pi(a) = c \, V_k \, S_q \otimes \mathbb{I}_2.
\end{equation*}
The associated state functionals are respectively:
\begin{align*}
    \omega_0(a) &= \braket{\zeta_0 | \pi(a) | \zeta_0} = c\braket{\zeta_0| V_k S_q \otimes \mathbb{I}_2|\zeta_0}, \\
    \omega_1(a) &= \braket{\zeta_1 | \pi(a) | \zeta_1} = c\braket{\zeta_1| V_k S_q \otimes 
    \mathbb{I}_2|\zeta_1}.
\end{align*}
The Weyl-restricted Connes distance between these two spatial states is therefore given by:
\begin{align}
    d_{W_\epsilon^B}(\omega_1, \omega_0) &= \sup_{k,q,c} \Big| \, \omega_1(a)  - \omega_0(a) \, \Big| \nonumber \\
    &= \sup_{k,q,c} \, \Big| \braket{\zeta_1 | \pi(a) | \zeta_1} - \braket{\zeta_0| \pi(a) |\zeta_0} \Big|,
\end{align}
subject to:
\begin{equation*}
    \| [W_\epsilon^B, \pi(a)] \| = c \left\| \sum_{\delta \in \mathcal{N}} S_\delta M_{K_\delta} \right\| \le \epsilon,
\end{equation*}
where $M_{K_\delta} = \sum_r \ket{r}\bra{r} \otimes K_\delta(r) $ and $K_\delta(r) = U_\delta(r + q) - e^{ik \cdot \delta} U_\delta(r)$.

To evaluate this constraint in the Landau gauge ($A_x = -By, A_y = 0$), we expand the local jump matrix:
\begin{align*}
    K_\delta(r) &= M_\epsilon H P_{s_y} H e^{-i \epsilon B(y+q_y) \sigma_z} P_{s_x}  \\
    &\quad - e^{ik\cdot \delta} M_\epsilon H P_{s_y} H e^{-i \epsilon By \sigma_z} P_{s_x}  \\
    &= M_\epsilon H P_{s_y} H e^{-i \epsilon By \sigma_z} \left( e^{-i \epsilon B q_y \sigma_z} - e^{ik\cdot \delta}\mathbb{I}_2 \right) P_{s_x}.
\end{align*}
The operator $P_{s_x}$ projects the coin space onto a single eigenspace characterized by the jump sign $s_x \in \{1,-1 \}$. Therefore, the diagonal phase difference matrix acting on this subspace reduces exactly to a scalar:
\begin{equation*}
    \left( e^{-i \epsilon B q_y \sigma_z} - e^{ik\cdot \delta}\mathbb{I}_2 \right) P_{s_x} = \left( e^{-is_x \epsilon B q_y} - e^{ik\cdot \delta} \right) P_{s_x}.
\end{equation*}
This makes $K_\delta(r)$ a rank one operator. Its norm is the absolute value of this scalar phase difference multiplied by the norm of the remaining coin operators. Since $\big\| M_\epsilon H P_{s_y} H P_{s_x} \big\| = 1/\sqrt{2}$, the norm of $K_\delta(r)$ is:
\begin{equation*}
    \| K_\delta \| = \frac{1}{\sqrt{2}} \left| 1 - e^{i(k_x s_x \epsilon + k_y s_y \epsilon + s_x \epsilon B q_y)} \right|.
\end{equation*}
The operator sum $\sum S_q M_{K_\delta}$ maps localized states to mutually orthogonal sites, its norm vanishes if and only if all four orthogonal jump branches vanish simultaneously. Setting the norm to zero requires satisfying the system of four equations:
\begin{equation} \label{4equations}
    s_x k_x \epsilon + s_y k_y \epsilon + s_x \epsilon B q_y = 0 \pmod{2\pi}, \quad \forall s_x, s_y \in \{\pm 1\}.
\end{equation}
For Eq. \ref{4equations} to hold independently of $s_x$ and $s_y$, the quantities $k_y \epsilon / \pi$ and $(k_x \epsilon + \epsilon B q_y) / \pi$ must be integers of the same parity. Focusing on the principal solution in the Brillouin zone $[-\pi/\epsilon, \pi/\epsilon]^2$, the y-momentum is equal to zero ($k_y = 0$) and the magnetic shift is balanced by the x-momentum ($k_x = -Bq_y$).

For these operators in the kernel, the state expectation value is proportional to $e^{-i B q_y r_x}$. Thus, if the two states are separated along the x-axis, the expectation difference is non-zero and can be scaled to infinity by $c \to \infty$, which implies:
\begin{equation}
    r_{0x} \neq r_{1x} \implies d_{W_\epsilon^B}(\omega_{r_1}, \omega_{r_0}) = \infty.
\end{equation}
This confirms that the metric disconnection observed in the continuum survives the lattice discretization.

\section{Dynamical Spectral Spread} \label{dynamical}
To characterize the geometric spread of the quantum walk, we consider two case studies. In the first, we study the 2+1D quantum walk defined in Sec. \ref{quantum-walk}. The Connes distance is evaluated between an initial Gaussian state and its time-evolved distribution in both continuous and discrete frameworks. In the absence of a magnetic field, we derive an analytical expression for the distance within the restricted family of position observables, corresponding to $q=0$ in the Weyl family, and verify the result numerically. We then study numerically the effect of a magnetic field associated with nonzero translation parameters $q$.

In the second case, we switch to the 1+1D plastic quantum walk introduced in \cite{dimolfetta2019quantumwalkcontinuoustimecontinuousspacetime}. Since the space is one dimensional, we use a localized initial state and the commutative algebra defined in Sec. \ref{comm-algebra}.
\subsection{2+1D Quantum Walk}
\subsubsection{Continuum Approximation}
The spread of the quantum walk is evaluated for the parametrized family of Weyl operators when $q=0$, $a = c \, T(k,0) = c \, e^{ik\cdot \hat{r}}$. The role of nonzero translation parameters $q$ is studied separately in a numerical analysis below. As shown in Sec.~\ref{noncomm-freedirac}, using the free Dirac operator, the constraint is simply $c \|k\| \le 1$. Scaling the test operator to saturate this constraint gives $c = 1/\|k\|$. 
This reduces the test observables to the commutative family of position observables whose representation is $\pi(a) = \frac{1}{\|k\|} e^{i k \cdot \hat{r}} \otimes \mathbb{I}_2$.

The algebraic states associated to the initial state $\psi_0$ and its time-evolution $\psi_t$ with distributions respectively $\rho_0(r)$ and $\rho_t(r)$ are:
\begin{align*}
    \omega_0 &= \frac{1}{\|k\|} \int_{\mathbb{R}^2} e^{i k \cdot r} \rho_0(r) d^2r, \\
    \omega_t &= \frac{1}{\|k\|} \int_{\mathbb{R}^2} e^{i k \cdot r} \rho_t(r) d^2r. 
\end{align*}

Therefore, the restricted Connes distance evaluated between $\omega_0$ and $\omega_t$ is:
\begin{equation}
    \begin{split}
    &d_{D_\text{free}}(\omega_t, \omega_0) \\
    &\quad = \sup_k \frac{1}{\|k\|}
      \left| \int_{\mathbb{R}^2} e^{i k \cdot r}
      \big[\rho_t(r)-\rho_0(r)\big] \, d^2r \right|.
    \end{split}
\end{equation}
The initial state is a Gaussian wavepacket centered at the origin with spatial width $\sigma$. Its probability density is $\rho_0(r) = \frac{1}{\pi\sigma^2} e^{-\frac{\|r\|^2}{\sigma^2}}$, and its algebraic state corresponds to its Fourier transform:
\begin{equation*}
   \frac{1}{\|k\|} \int_{\mathbb{R}^2} e^{i k \cdot r} \rho_0(r) d^2r = \frac{1}{\|k\|} e^{-\frac{\sigma^2 \|k\|^2}{4}}.
\end{equation*}
In the absence of a magnetic field, the wavepacket expands ballistically. To obtain an analytical estimate of the macroscopic spread, we approximate the free propagating probability profile by a ballistic ring whose radius grows at the maximum causal speed, $R(t) = t$. The density of this thin ring is approximated using a radial Dirac delta $\frac{1}{2\pi R} \delta(r - R)$ in polar coordinates $(r, \theta)$. Its Fourier transform evaluates to a Bessel function of the first kind of order zero (see Appendix \ref{bessel-proof}):
\begin{equation*}
    \int_{\mathbb{R}^2} e^{i k \cdot r} \frac{1}{2\pi R} \delta(r - R) \, d^2r = J_0(\|k\|R).
\end{equation*}
We also approximate the evolved probability density $\rho_t(r)$ as the spatial convolution of the initial Gaussian density $\rho_0(r)$ with this ideal ballistic ring. Under this approximation, the convolution theorem allows us to express the Fourier transform of the probability density as the product of their individual Fourier transforms:
\begin{equation*}
    \int_{\mathbb{R}^2} e^{i k \cdot r} \rho_t(r) d^2r = e^{-\frac{\sigma^2 \|k\|^2}{4}} J_0(\|k\|R).
\end{equation*}
Thus, the restricted Connes distance is:
\begin{align} \label{discrete-connes-dist-spread}
    d_{D_\text{free}}(\omega_t, \omega_0) &= \sup_k \frac{1}{\|k\|} \left\| e^{-\frac{\sigma^2 \|k\|^2}{4}} J_0(\|k\|R) - e^{-\frac{\sigma^2 \|k\|^2}{4}} \right\| \nonumber \\
    &= \sup_k \left[ e^{-\frac{\sigma^2 \|k\|^2}{4}} \frac{1 - J_0(\|k\|R)}{\|k\|} \right].
\end{align}
We introduce the dimensionless variables $\alpha = \|k\|R$ and $\gamma = \frac{\sigma}{2R}$. The distance can be rewritten as $d_{D_\text{free}}(\omega_t, \omega_0) = R \cdot \sup_\alpha g(\alpha)$, where the objective function to maximize is:
\begin{equation*}
    g(\alpha) = e^{-\gamma^2\alpha^2} \frac{1 - J_0(\alpha)}{\alpha}.
\end{equation*}
Setting the derivative $g'(\alpha) = 0$ gives the extremum condition:
\begin{equation*}
    (2\gamma^2\alpha^2 + 1)(1 - J_0(\alpha)) = \alpha J_1(\alpha).
\end{equation*}
There are two distinct physical regimes depending on the evolution time:
\begin{itemize}
    \item $t \to \infty$: As the quantum walk evolves over long durations, the radius $R$ dominates the initial Gaussian width $\sigma$, making the ratio $\gamma \to 0$. The extremum condition simplifies to $1 - J_0(\alpha) = \alpha J_1(\alpha)$, which is numerically satisfied at $\alpha_{\max} \approx 2.75826$. Evaluating the objective function at this peak gives $\sup_\alpha g(\alpha) \approx 0.423$. Thus, the restricted Connes distance grows linearly with time:
    \begin{equation}
        d_{D_\text{free}}(\omega_t, \omega_0) \approx 0.423 R = 0.423 \cdot t.
    \end{equation}

    \item $t \to 0$: In the early stages of the walk, the initial Gaussian width dictates the behavior, meaning $\gamma \gg 1$. The maximum is reached at small values of $\alpha$ because of the strong exponential decay, which allows us to substitute the first-order Taylor expansion $J_0(\alpha) \approx 1 - \frac{\alpha^2}{4}$. The objective function simplifies to $g(\alpha) \approx \frac{\alpha}{4} e^{-\gamma^2\alpha^2}$. Setting the derivative to zero gives the maximum at $\alpha_{\max} = \frac{1}{\sqrt{2}\gamma} = \frac{\sqrt{2}R}{\sigma}$. Substituting this back into the distance formula gives a quadratic early-time expansion:
    \begin{equation}
        d_{D_\text{free}}(\omega_t, \omega_0) \approx R \left( \frac{R}{2\sqrt{2}\sigma\sqrt{e}} \right) = \frac{ t^2}{2\sqrt{2}\sigma\sqrt{e}}.
    \end{equation}
\end{itemize}

\subsubsection{Discrete Result} \label{discrete-comm-algebra}
In the discrete case, the representation of the reduced commutative family of position observables is $\pi(a) = c V_k \otimes \mathbb{I}_2$. In the absence of a magnetic field, the metric constraint is given by:
\begin{equation*}
    \| [W_\epsilon, \pi(a)] \| = c \left\|\sum_{\delta \in \mathcal{N}} S_\delta \otimes K_\delta \right\| \le \epsilon,
\end{equation*}
where the matrix $K_\delta(r)$ loses its position dependence and becomes $K_\delta = U_\delta (1 - e^{i k \cdot \delta})$, with $U_\delta = M_\epsilon H P_{s_y} H P_{s_x}$.
Since the operator $\sum_{\delta} S_\delta \otimes K_\delta$ is translation-invariant, we evaluate its norm by diagonalizing it in the Fourier domain.
Let $W_\epsilon(p)$ denote the $2 \times 2$ walk operator in momentum space. In this domain, the position observable $V_k = e^{ik \cdot \hat{r}}$ shifts the quasimomentum, which gives the following relation for the commutator:
\begin{equation*}
    \mathcal{F}[W_\epsilon, V_k]\mathcal{F}^{-1} = W_\epsilon(p-k) - W_\epsilon(p),
\end{equation*}
where $\mathcal{F}$ denotes the Fourier transform.
The operator norm of the commutator is therefore determined by the supremum over all quasimomenta $p$:
\begin{equation*}
    \|[W_\epsilon, V_k]\| = \sup_p \|W_\epsilon(p-k) - W_\epsilon(p)\|.
\end{equation*}
For small momentum $k$, applying a Taylor expansion gives: 
\begin{equation*}
    \|[W_\epsilon, V_k]\| = \|k\| \sup_p \left\| \hat{k} \cdot \nabla_p W_\epsilon(p) \right\| + \mathcal{O}(\|k\|^2).
\end{equation*}
For the free Dirac walk, the supremum of the group velocity $\sup_p \|\hat{k} \cdot \nabla_p W_\epsilon(p)\|$ equals $1$. Thus, the discrete commutator norm becomes $\|[W_\epsilon, \pi(a)]\| \approx c \|k\|$, reducing the metric constraint $\|[W_\epsilon, \pi(a)]\| \le \epsilon$ to a scaling factor $c = \frac{\epsilon}{\|k\|}$.

The expectation value of this saturated test operator on the evolved state is $\braket{\psi_N | \pi(a) | \psi_N } = \frac{\epsilon}{\|k\|} \tilde{\rho}_N(k)$, where $\tilde{\rho}_N(k)$ is the Fourier transform of the position probability density of the walker. Therefore, the restricted discrete Connes distance is:
\begin{equation*}
    d_{W_\epsilon}(\omega_N, \omega_0) = \epsilon \cdot \sup_k \frac{|\tilde{\rho}_N(k) - \tilde{\rho}_0(k)|}{\|k\|}
\end{equation*}
We consider, as we did in the continuous setting, an initial discrete Gaussian wavepacket localized at the origin with spatial width $\sigma$:
\begin{equation*}
    \rho_0(r) = \frac{1}{C^2} e^{-\frac{\|r\|^2}{\sigma^2}},
\end{equation*}  
where $C$ is the normalization constant. Its Fourier transform in momentum space is $\tilde{\rho}_0(k) = e^{-\frac{\sigma^2 \|k\|^2}{4}}$.

Translation invariance implies that the amplitude of the evolved state $\psi_N(r)$ is a convolution of the initial amplitude with the discrete propagator. However, the resulting probability distribution $|\psi_N(r)|^2$ contains interference cross-terms. In the long-wavelength regime, we approximate the discrete propagator by its continuum ballistic profile $\tilde{\rho}_{pt}(k) \approx J_0(\|k\|N)$. Within this approximation, the evolved probability density is:
\begin{equation*}
    \tilde{\rho}_N(k) \simeq e^{-\frac{\sigma^2 \|k\|^2}{4}} J_0(\|k\| N).
\end{equation*}
Substituting this back gives the restricted Connes distance:
\begin{equation}
    d_{W_\epsilon}(\omega_N, \omega_0) = \epsilon \cdot \sup_k \left[ e^{-\frac{\sigma^2 \|k\|^2}{4}} \frac{1 - J_0(\|k\|N)}{\|k\|} \right].
\end{equation}
We recover the analogue result of the continuous setting (Eq. \ref{discrete-connes-dist-spread}), leading to the same analysis and conclusion for the two time regimes.
In the following section, we compute the restricted Connes distance numerically in the absence of the magnetic field and for $q=0$, and we compare it to the obtained results. Then, we analyse the result of the implementation when $q \neq 0 $ in a magnetic field.

\subsubsection{Numerical Analysis}

We now evaluate the Connes distance directly on a finite periodic $L \times L$ lattice, without imposing the continuum approximations used in the previous sections. This provides an independent numerical test of the analytical predictions. In our 2+1D simulations, we use a $64 \times 64$ lattice with grid spacing and time step $\epsilon = 0.01$, mass $m = 1.0$, and an initial Gaussian width of $\sigma \approx 2.26\epsilon$ (further numerical details are provided in Appendix \ref{numerical-details}). All vectors $k$ in the discrete evaluation are expressed in dimensionless lattice units ($k_{\mathrm{lat}} = \epsilon k_{\mathrm{phys}}$), which scales the distance by a factor of $\epsilon$. In a similar way, we distinguish the physical magnetic field $B_{\mathrm{phys}}$ from the dimensionless plaquette flux $\Phi = B_{\mathrm{phys}}\epsilon^2$.

\paragraph{Free Evolution}

\begin{figure*}[htbp]
    \centering
    \includegraphics[width=\textwidth]{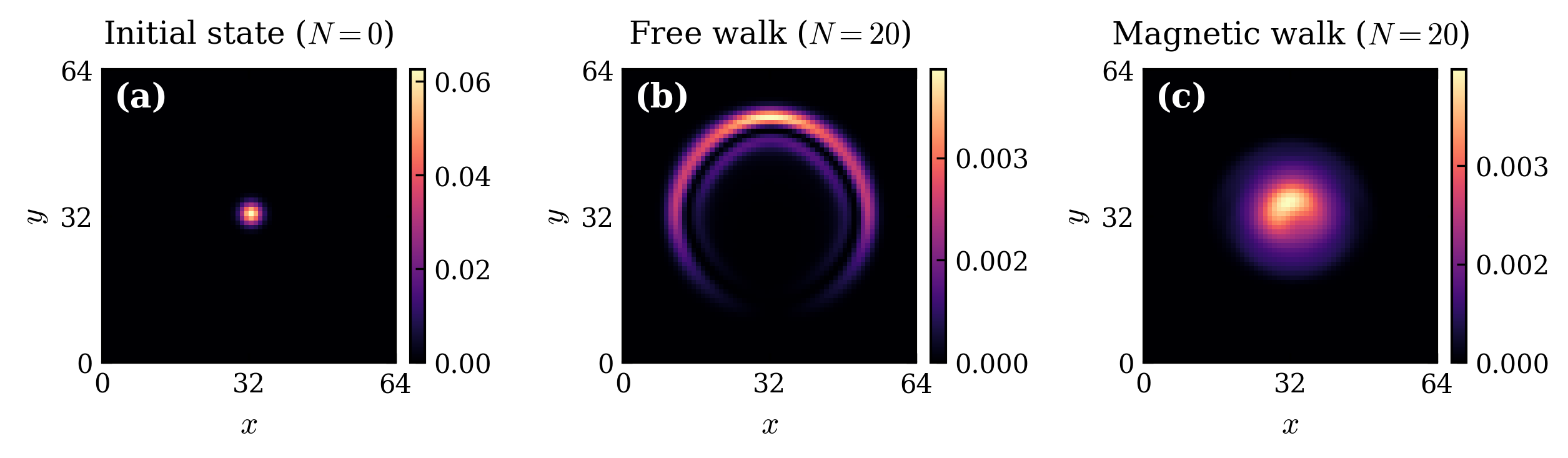}
    \caption{Wavepacket probability densities for the discrete quantum walk. (a) Initial localized Gaussian state at $t=0$. (b) Free walk evolution at $N=20$ showing the macroscopic ballistic ring. (c) Magnetic walk evolution at $N=20$ demonstrating spatial suppression and Landau-like confinement.}
    \label{fig:densities}
\end{figure*}

We first consider the free walk ($B=0$) restricted to the commutative subfamily of position observables ($q=0$). Figure \ref{fig:densities} illustrates the real-space probability density of the walker after $N=20$ time steps. The free evolution shows a clear ballistic ring structure, providing qualitative support for the thin-ring approximation used in our analytical derivations. 

The resulting numerical distance $d_{\rm num}(N)$ is compared with the analytical ballistic prediction $d_{W_\epsilon}(N)$, which we denote as $d_\text{ring} = 0.423 N \epsilon$ for the remainder of this section. The coefficient $0.423$ is obtained under the idealized assumption that the probability distribution is concentrated on an infinitely thin ring at the corresponding ballistic radius. However, the numerical walk does not satisfy this assumption because the finite-width wavepacket contains interference fringes and a nonzero distribution away from the ring. Therefore, the thin-ring estimate captures the leading geometric scale but is not expected to coincide exactly with the numerical Connes distance.

\begin{figure}[htbp]
    \centering
    \includegraphics[width=\columnwidth]{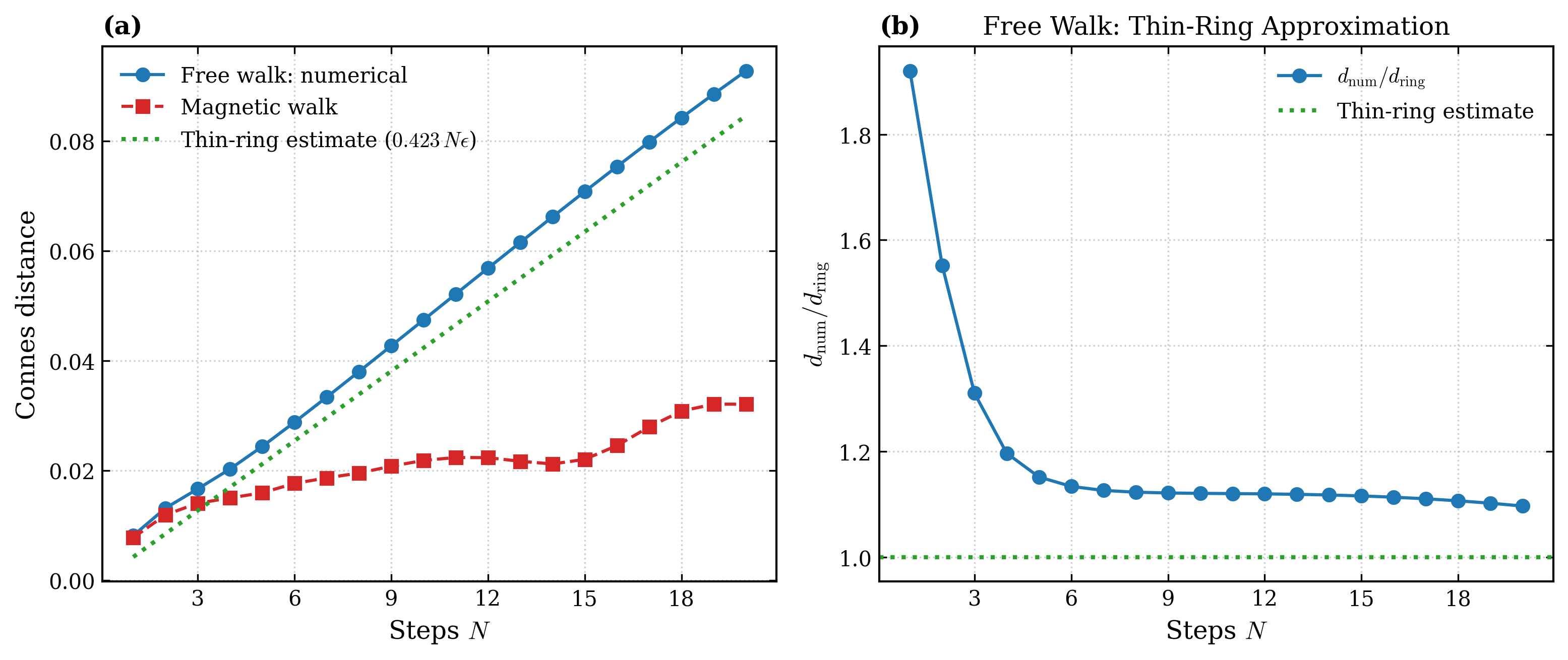}
    \caption{(a) Numerical evaluation of the Connes distance for the free and magnetic quantum walks compared to their theoretical limits. (b) The numerical-to-asymptotic ratio $\mathcal{R}(N)$ for the free walk, which rapidly stabilizes near $1.12$.}
    \label{fig:connes_distance}
\end{figure}

As shown in Fig.~\ref{fig:connes_distance}(a), following a brief initial transition phase, the macroscopic spread converges into a linear ballistic expansion. It remains systematically above the thin-ring estimate which is not a numerical error but a consequence of the idealized ring approximation.

To quantify the deviation between the exact lattice dynamics and the continuum theory, we introduce the ratio
\begin{equation*}
    \mathcal{R}(N) = \frac{d_{ num}}{d_\text{ring}}.
\end{equation*}
As illustrated in Fig. \ref{fig:connes_distance}(b), this ratio approaches a slowly varying value $\mathcal{R}=1.12$ rather than converging to one, further demonstrating that the thin-ring expression should be interpreted as a geometric reference estimate rather than as the exact asymptotic coefficient.

The numerical results therefore confirm the expected ballistic scaling of the free walk while also quantifying the limitations of the thin-ring approximation. In particular, the linear dependence on $N$ is robust, whereas the numerical prefactor depends on the detailed spatial distribution and interference structure of the wavepacket. This distinction becomes important when comparing the free walk with the magnetic walk, for which the spatially varying phases modify the propagation and can lead to a different Connes distance.

\paragraph{Magnetic Evolution}

Next, we introduce a uniform magnetic field corresponding to a dimensionless plaquette flux $\Phi = B_{\mathrm{phys}}\epsilon^2 = 2\pi/L$ through the Landau gauge $\vec{A} = (-By, 0)$. In the presence of the magnetic field, translation invariance is modified, and we therefore enlarge the family of test operators to include spatial translations,
\begin{equation*}
    a_{k,q}=V_k T_q, \qquad q=(0,q_y).
\end{equation*}
While the distance diverges between two spatially separated states, the dynamical spread $d_{W_\epsilon}(\omega_N, \omega_0)$ is bounded by the causal light cone ($N\epsilon$), which is obtained by telescoping the unitary evolution under the commutator constraint. Kernel observables ($[W_\epsilon^B, \pi(a)] = 0$) correspond to conserved symmetries whose expectation values remain static ($\omega_N(a) - \omega_0(a) = 0$). To capture the geometric expansion rather than these static symmetries, we define a non-degenerate distance by excluding these operators from the numerical optimization ($\left\|[W_\epsilon^B,V_k T_q]\right\|>0$). This exclusion removes a $0/0$ from the ratio formula, it does not solve an infinite dynamical distance, as the spread is inherently finite. As shown in Fig. \ref{fig:densities}(c), the magnetic walk has a reduced spatial spread compared with the free walk, consistent with a Landau-like confinement induced by the magnetic field. The same behavior is reflected by the non-degenerate distance in Fig. \ref{fig:connes_distance}(a): after an initial approximative linear growth, the distance progressively saturates towards a plateau. Thus, the magnetic field suppresses the long-time growth of the wavepacket's spread.

\subsection{1+1D Plastic Quantum Walk}
As a second case study, we evaluate the spread of the 1+1D plastic quantum walk, which simulates a Dirac particle in an inhomogeneous spatial metric \cite{dimolfetta2019quantumwalkcontinuoustimecontinuousspacetime}. The discrete evolution operator for this system is given by
\begin{equation}
    W_\epsilon = \Lambda^{-1} S C_{-\zeta} S C_\zeta \Lambda
\end{equation}
where $S$ is the spin-dependent spatial shift, $C_{\pm\zeta}$ are coin operations parameterized by a mixing angle $\theta$, and $\Lambda$ is a position-dependent unitary operator encoding the spacetime metric. 
They are expressed in the standard computational basis by:
\begin{equation*}
    C_\zeta = 
    \begin{pmatrix} 
    -\cos\theta & e^{-i\zeta}\sin\theta \\ 
    e^{i\zeta}\sin\theta & \cos\theta 
    \end{pmatrix}, 
    \quad 
    \Lambda = \frac{1}{2}
    \begin{pmatrix} 
    -f_- & f_+ \\ 
    f_+ & f_- 
    \end{pmatrix},
\end{equation*}
where $f_\pm(x) = \sqrt{1-c(x)} \pm \sqrt{1+c(x)}$ and $c(x)$ is the local speed of light.

The parameter $\zeta$ acts as a mass and phase parameter, and $\theta$ controls the propagation speed. Throughout this section, we restrict the walk to the massless regime ($\zeta = 0$) and the coin parameter range $0 < \theta < \pi/2$.

We first analyze the homogeneous case where $\theta$ is a constant, and we express the Connes distance using Konno's weak limit theorem \cite{konno2008quantum}. We then switch to numerical simulations to study the wavepacket propagation in curved geometries ($\theta = \theta(x)$).

\subsubsection{The Spread in the Homogeneous Limit}
On the finite lattice used later for the numerical evaluation, we restrict the test observables to the commutative $C^*$-algebra of diagonal spatial observables represented by $\pi(a) = \sum_x a_x \ket{x}\bra{x} \otimes \mathbb I_2$. In this framework, the discrete Dirac operator is defined as $D \approx \frac{i}{2\epsilon}(W_\epsilon - I)$, which modifies the constraint of the Connes distance to $\| [W_\epsilon, \pi(a)] \| \le 2\epsilon$.

Since the evolution operator contains two sequential shift operators $S$, a single time step displaces the walker by $\delta \in \{-2, 0, 2\}$. The commutator action on a localized state gives a dependence on the discrete spatial gradient $\Delta_x = a_{x+2} - a_x$. The operator norm for the commutator is: (see Appendix \ref{plastic-qw}):
\begin{equation}
    \|[W_\epsilon, \pi(a)]\| = |\cos \theta| \max_x |\Delta_x|.
\end{equation}
Saturating the constraint $\|[W_\epsilon, \pi(a)]\| \le 2\epsilon$ bounds the discrete spatial gradient:
\begin{equation}
    \max_x |\Delta_x| \leq \frac{2\epsilon}{|\cos\theta|}.
\end{equation}
We sum this gradient constraint over a path of length $|x|/2$ jumps, giving:
\begin{equation*}
    |a_x - a_0| \le |x| \frac{\epsilon}{|\cos \theta|}.
\end{equation*}

We compute the Connes distance between an initial localized state at the origin $\ket{\psi_0} = \ket{0} \otimes \ket{c_0}$ and the evolved state $\ket{\psi_t} = W_\epsilon^N \ket{\psi_0}$. 
The associated state functionals are respectively:
\begin{align*}
    \omega_0(a) &= \braket{\psi_0 | \pi(a) | \psi_0} = a_0, \\
    \omega_t(a) &= \braket{\psi_t | \pi(a) | \psi_t} = \sum_{x} a_x P_t(x),
\end{align*}
where $P_t(x) = \sum_c |\psi_t(x, c)|^2$ is the spatial probability distribution at time $t$.

The Connes distance is obtained by optimizing the expectation value difference over all admissible observables. In the present homogeneous setting, the translation-invariant constraint implies $|a_{x+2}-a_x| \leq \frac{2\epsilon}{|\cos\theta|}$. The walk remains on the even sublattice, and an extremal observable can be chosen as
\begin{equation*}
    a_x=a_0+\frac{\epsilon}{|\cos\theta|}|x|, \qquad x\in2\mathbb Z,
\end{equation*}
for which $|a_{x+2}-a_x| =\frac{2\epsilon}{|\cos\theta|}$ on the even sublattice. 
(Note: On an infinite lattice, we use bounded truncation of this observable outside the causal support of the walk. On a finite periodic lattice, $|x|$ represents the shortest path graph distance avoiding the periodic wrap.)

Thus, the constraint can be saturated simultaneously over the support of the walk. The Connes distance reduces to:
\begin{align}
    d_{W_\epsilon}(\omega_t, \omega_0) &= \sup_a \left| \omega_t(a) - \omega_0(a) \right| \nonumber \\
    &= \sup_a \left| \sum_{x} a_x P_t(x)- a_0 \sum_x P_t(x) \right| \nonumber\\
    &= \sup_a \left| \sum_{x} (a_x - a_0) P_t(x) \right| \nonumber\\
    &= \frac{\epsilon}{|\cos\theta|} \sum_x |x|P_t(x) \nonumber\\
    &= \frac{\epsilon}{|\cos \theta|} \braket{|x|}_t.
\end{align}

To obtain an analytical expression for the asymptotic macroscopic spread ($t \to \infty$), we apply Konno's weak limit theorem \cite{konno2008quantum}. For an initial coin state $\ket{c_0}$, the rescaled position variable $v = x / (2N)$ converges in distribution to the continuous probability density function $f(v)$. Generally, $f(v)$ consists of a symmetric Konno factor multiplied by a linear state-dependent asymmetric factor \cite{konno2005newtypelimittheorems}. However, since the Connes distance evaluates the even test function $\braket{|v|}$, the contribution from the state-dependent odd component vanishes by parity.
Therefore, the asymptotic spread is independent of the initial coin state and is evaluated using only the symmetric part of the limit density:
\begin{equation*}
    f_\text{sym}(v) = \frac{\sin \theta}{\pi(1 - v^2)\sqrt{\cos^2 \theta - v^2}} \quad \text{for} \quad |v| < \cos \theta.
\end{equation*}
Therefore, the expected value of the absolute position is:
\begin{align}
    \braket{|x|}_t \approx 2N \int_{-\cos \theta}^{\cos \theta} |v| f_\text{sym}(v) dv &= 4N \int_{0}^{\cos \theta} v f_\text{sym}(v) dv \nonumber\\
    &= \frac{4N}{\pi} \sin \theta \left( \frac{\frac{\pi}{2} - \theta}{\sin \theta} \right) \nonumber\\
    &= \frac{t}{\epsilon} \left( 1 - \frac{2\theta}{\pi} \right), \nonumber
\end{align}
where $t= 2 N \epsilon$.
Substituting this back into the distance, we obtain the asymptotic Connes distance for the 1+1D plastic quantum walk:
\begin{equation}
    d_{W_\epsilon}(\omega_t, \omega_0) \approx \frac{t}{|\cos \theta|} \left( 1 - \frac{2\theta}{\pi} \right).
\end{equation}
This confirms that the spectral spread of the homogeneous plastic walk grows linearly with time.

\begin{figure*}[htbp]
    \centering
    \includegraphics[width=\textwidth]{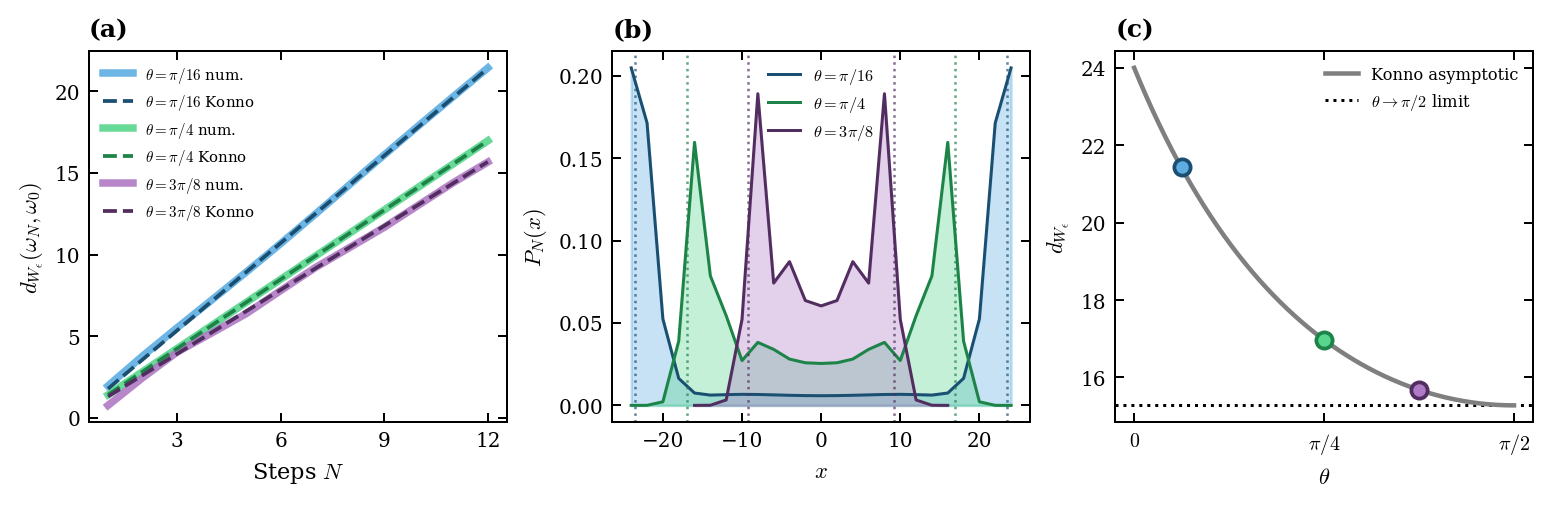}
    \caption{Spread of the homogeneous 1+1D plastic quantum walk. (a) The numerical Connes distance plotted against walk steps $N$ for different values of $\theta$. (b) The spatial probability density $P_N(x)$ at $N=12$ for different values of $\theta$. The dotted vertical lines are the causal edges $x = \pm 2N\epsilon\cos\theta$. (c) Connes distance as a function of the coin parameter $\theta$ at fixed $N$, comparing the numerical values (colored markers) with the analytical asymptotic prediction.}
    \label{fig:plastic_homogeneous}
\end{figure*}

Figure \ref{fig:plastic_homogeneous} provides a numerical validation of the analytical result. For a finite homogeneous lattice, we independently evaluate the Connes distance supremum by convex optimization.

At $N=8$ and $\theta = \pi / 4$, the optimized distance agrees with the analytic expression, and the optimized operator saturates the constraint to numerical precision.

As shown in Fig. \ref{fig:plastic_homogeneous}(a), the numerical Connes distance is consistent with the linear ballistic growth, matching the Konno asymptotic prediction for large $N$. The corresponding probability densities in Fig. \ref{fig:plastic_homogeneous}(b) are macroscopically confined within the asymptotic causal edges $x = \pm 2N\epsilon\cos\theta$ with only finite quantum tails extending beyond the asymptotic light cone. 
The dependence on the coin parameter is further tested in Fig. \ref{fig:plastic_homogeneous}(c), the numerical values closely agree with the analytical curve and its $\theta$-dependence, which combines the geometric factor $1/|\cos\theta|$ from the constraint and the $\theta$-dependent ballistic prefactor from the Konno distribution.

\subsubsection{Numerical Simulation of an Inhomogeneous Spatial Metric}
To simulate an inhomogeneous spatial metric, the original formulation of the plastic walk \cite{dimolfetta2019quantumwalkcontinuoustimecontinuousspacetime}\cite{PhysRevA.88.042301} allows the propagation speed to vary spatially. We consider the Lorentzian spacetime metric 
\begin{equation*}
    ds^2 = dt^2 - c(x)^{-2}dx^2,
\end{equation*}
whose spatial component is $g_{11}(x) = -c(x)^{-2}$. The induced positive spatial metric is therefore $h_{11}(x) = c(x)^{-2}$, giving the spatial line element 
\begin{equation*}
    dl = \frac{|dx|}{c(x)}.
\end{equation*}
Following the inhomogeneous quantum walk construction \cite{Arrighi2018quantumwalkingin}, we encode the propagation speed through a position-dependent coin parameter 
\begin{equation*}
    \cos \theta(x) = \kappa c(x),
\end{equation*}
so that the coin operators become spatially dependent.

Introducing $C_{\theta(x)}$ breaks translation invariance and changes the local structure of the lattice.
In flat spacetime, the translation-invariant constraint simplifies the Connes distance to a function of $\braket{|x|}$. This reduction no longer holds for a spatially changing $c(x)$. Therefore, the numerical distance is computed directly from its defining expression: 
\begin{equation}
    d_{W_\epsilon}(\omega_N,\omega_0) = \sup_a \left\{ \omega_N(a)-\omega_0(a) : \|[W_\epsilon,\pi(a)]\| \le 2\epsilon \right\}.
\end{equation}
In general, when $c = c(x)$, the homogeneous relation $d_{W_\epsilon} \propto \braket{|x|}_N$ is no longer valid.

\begin{figure*}[htbp]
    \centering
    \includegraphics[width=\textwidth]{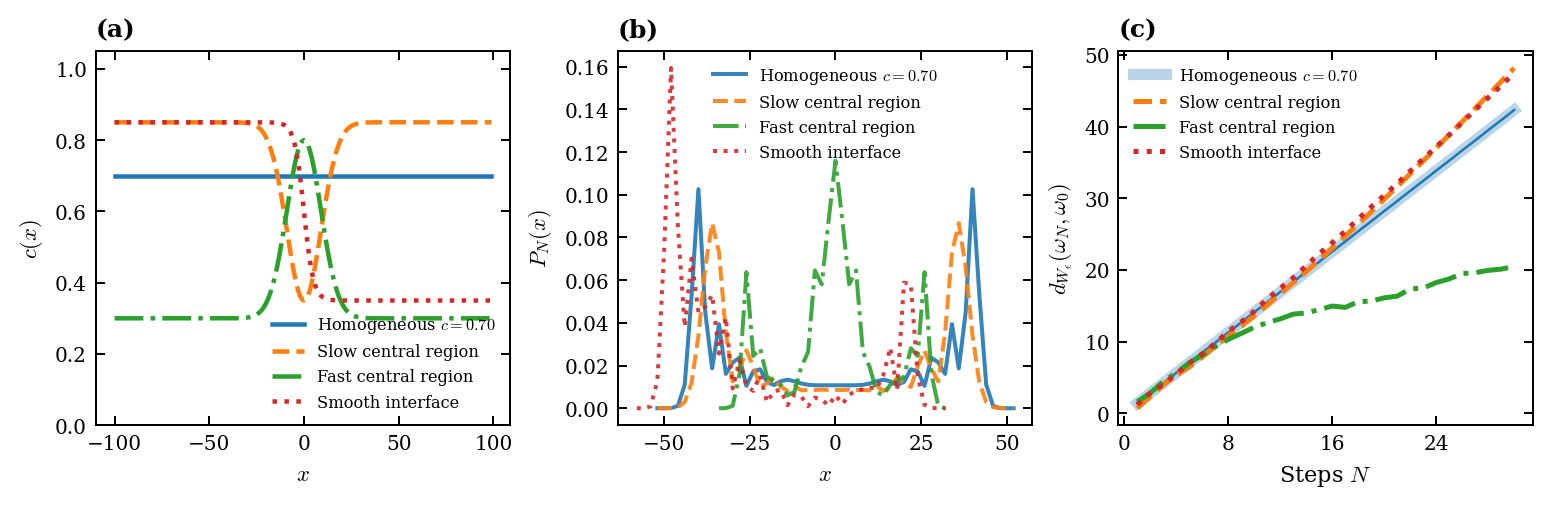}
    \caption{Numerical evaluation of the Connes distance in an inhomogeneous spatial metric. (a) Spatial profiles of the speed of light $c(x)=\cos\theta(x)$ (setting $\kappa=1$), which encode the simulated metric $g_{11}$. (b) Distorted wavepacket probability densities $P_N(x)$ at $N=30$. (c) The numerical Connes distance plotted against walk steps $N$.}
    \label{fig:plastic_curved}
\end{figure*}

We then compare several profiles of the speed of light in Fig.~\ref{fig:plastic_curved}(a) to grasp the hidden physical phenomena. Specifically, we compare a flat homogeneous baseline (solid blue line) against three non-homogeneous geometries: a slow central region where the propagation speed symmetrically dips at the origin (dashed orange curve), a fast central region where the speed peaks at the origin (dash-dot green curve), and a smooth interface representing a continuous step-like transition from a high speed on the left to a low speed on the right (dotted red curve).

As shown in Fig. \ref{fig:plastic_curved}(b), spatial variations in the speed of light $c(x)$ heavily distort wavepacket propagation. A fast central region surrounded by slower boundaries (dash-dot green curve) induces strong spatial gradients. This variation produces reflection and interference, which leads to a central accumulation at the origin. In the opposite case, a slow central region (dashed orange curve) initially suppresses propagation near the origin, after which the wavepacket spreads more rapidly upon entering the faster outer regions.

These dynamics are evaluated using the numerical Connes distance and plotted in Fig.~\ref{fig:plastic_curved}(c). Unlike the linear ballistic growth observed in the homogeneous case, the distance in the non-homogeneous geometries is non-linear in time. It reflects both the spatial probability distribution and the position-dependent constraint. Regions with larger speed of light $c(x)$ impose tighter restrictions on the gradient, whereas regions with a smaller $c(x)$ relax this constraint, allowing larger admissible spatial gradients and generating a larger geometric distance.

Therefore, the same coordinate-space displacement can contribute differently to the Connes distance depending on where the wavepacket propagates. For instance, while the green profile produces a central accumulation in a fast region, resulting in a highly suppressed Connes distance, wavepackets, spending time in a slow region such as the center of the orange curve or the right-hand side of the smooth interface (dotted red curve), experience a locally stretched spatial metric. Thus, even when the wavepacket decelerates in coordinate space, its geometric distance grows rapidly, surpassing the homogeneous space baseline.

This behavior follows from the spatially dependent constraint. In the continuum limit, the local geometric cost is consistent with the spatial line element
\begin{equation*}
    dl=\frac{|dx|}{c(x)}.
\end{equation*}
The Connes distance therefore provides a geometry-sensitive measure that remains well defined when the homogeneous reduction breaks down, and adapts to spatial variations in the metric.

\section{Conclusion}
In this work, we developed a discrete spectral geometry framework to evaluate the dynamical spread of quantum walks using the Connes distance. By replacing the continuous Dirac operator with the unitary step operator, we successfully mapped continuous metric constraints onto lattice dynamics. In the $2+1$D Dirac walk, we showed that the underlying geometry is dictated by the choice of observable algebra. While a commutative position algebra remains blind to magnetic fields, expanding to the noncommutative family of Weyl operators captures the gauge-momentum coupling. This coupling leads to a nontrivial kernel in the metric constraint, which causes a breakdown of the spectral metric along specific algebraic directions.

Furthermore, we proved the robustness of this discrete metric in an inhomogeneous settings through the 1+1D plastic quantum walk \cite{dimolfetta2019quantumwalkcontinuoustimecontinuousspacetime}. We showed that the Connes distance adapts to spatially varying propagation speeds, capturing the local metric costs of simulated inhomogeneous spatial metrics. 
Setting the coin mixing angle to $\theta = \pi/4$ in the massless regime recovers a standard balanced quantum walk. While its specific coin operator differs from the canonical Hadamard matrix by relative phases, it belongs to the same universality class and shares identical macroscopic expansion properties.
Our analytical results, grounded in Konno's limit theorem, therefore might directly apply to the widely studied Hadamard walk.

Looking forward, a natural extension of this work would be to evaluate the Connes distance over the full algebra of Weyl operators, rather than a restricted family. It would also be interesting to explore whether the present discrete framework can be extended to quantum walks on intrinsically noncommutative geometries, such as the fuzzy sphere. Beyond these mathematical extensions, the discrete Connes distance and the corresponding Python numerical framework developed in this work offer a powerful mathematical tool with broad applicability in quantum information and quantum transport. In the context of quantum transport, because the Connes metric restricts how fast a test observable can vary across the lattice, it provides a natural way to quantify transport efficiency, the trapping of wavepackets in disordered lattices, or the propagation of waves along boundaries in topological materials.
In quantum information theory, state distinguishability is usually measured by metrics like the trace distance or fidelity, which are blind to the underlying spatial structure of the system \cite{Trevisan2025, De_Palma_2021}. The Connes distance, by contrast, provides a geometry-aware measure of distinguishability. It offers a geometric approach to deriving Lieb-Robinson-type bounds for information propagation~\cite{Bravyi_2006}.

\section*{Data and Code Availability}
The numerical framework developed for this study is implemented in Python. The source code used to generate the data and figures in this paper is available at \url{https://gitlab.inria.fr/parrighi/connes-qw}.

\begin{acknowledgments}
\RaggedRight
%
%
This work was supported by the Institute Quantum-Saclay and the QuanTEdu-France program. Partial funding was provided by the European Union through the MSCA SE project QCOMICAL. We also acknowledge support from the French National Research Agency (ANR) through the TaQC project (Grant No. ANR-22-CE47-0012) and the ``Plan France 2030'' framework under grants EPIQ (ANR-22-PETQ-0007), OQULUS (ANR-23-PETQ-0013), HQI-Acquisition (ANR-22-PNCQ-0001), and HQI-R\&D (ANR-22-PNCQ-0002).

Additional support was provided by the WithOut SpaceTime (WOST) project (\url{https://withoutspacetime.org}) under Grant No. 63683 from the John Templeton Foundation (JTF). The opinions expressed in this publication are those of the authors and do not necessarily reflect the views of the John Templeton Foundation.

All three authors acknowledge funding from the UE-Colombian Stic-Amsud QUASAR project (Grant No. 250016). A.~F. Reyes-Lega acknowledges financial support from the Vice Presidency of Research and Creation at Universidad de los Andes (Project No. INV-2025-220-3480) and the Faculty of Sciences of Universidad de los Andes (Project No. INV-2026-244-4244).

\end{acknowledgments}

\clearpage
\appendix

\section{Continuous Framework Proofs}

\subsection{Commutative Limit and Euclidean Distance} \label{euclidean1}
Let $\ket{\zeta_1}$ and $\ket{\zeta_2}$ be two normalized Gaussian states centered respectively at $r_1$ and $r_2$ with the same width $\sigma$. The algebraic states evaluated on a spatial observable $f(\hat{r})$ are given by their expectation values:
\begin{equation*}
    \omega_{r_i}(f) = \braket{\zeta_i| f(\hat{r}) |\zeta_i} = \int_{\mathbb{R}^2} f(r) \rho_i(r) d^2r,
\end{equation*}
where $\rho_i(r) = |\zeta_i(r)|^2$. Since the states share the same width, we can define a common centered density $\rho_0(r)$ such that $\rho_i(r) = \rho_0(r - r_i)$.

Therefore, the Connes distance is:
\begin{equation*}
    \begin{split}
    &d_D(\omega_{r_1}, \omega_{r_2}) \\
    &\quad = \sup_{\|\nabla f\| \le 1}
      \left| \int_{\mathbb{R}^2} f(r)
      \big[\rho_1(r)-\rho_2(r)\big] \, d^2r \right|.
    \end{split}
\end{equation*}
Changing variables to $x = r - r_i$ in each respective integral gives:
\begin{equation*}
    \begin{split}
    &d_D(\omega_{r_1}, \omega_{r_2}) \\
    &\quad = \sup_{\|\nabla f\| \le 1}
      \left| \int_{\mathbb{R}^2}
      \big[f(x+r_1)-f(x+r_2)\big] \rho_0(x) \, d^2x \right|.
    \end{split}
\end{equation*}
The constraint $\|\nabla f\| \le 1$ implies $|f(x + r_1) - f(x + r_2)| \le \|r_1 - r_2\|$. Since $\rho_0(x)$ is a normalized probability distribution, this directly gives the upper bound:
\begin{equation*}
    d_D(\omega_{r_1}, \omega_{r_2}) \le \int_{\mathbb{R}^2} \|r_1 - r_2\| \rho_0(x) \, d^2 x = \|r_1 - r_2\|.
\end{equation*}

On the other hand, to establish the lower bound, we saturate the constraint by choosing a bounded, $1$-Lipschitz truncation $f_R(r)$ that matches the optimal linear function $f(r) = \frac{r_1 - r_2}{\|r_1 - r_2\|} \cdot r$ inside a ball of radius $R$, and is constant outside. Taking the limit $R \to \infty$, the integral becomes:
\begin{equation*}
    \int_{\mathbb{R}^2} f(r) \rho_i(r) d^2r = \int_{\mathbb{R}^2} \big[ f(x) + f(r_i) \big] \rho_0(x) d^2x.
\end{equation*}
Since the Gaussian density $\rho_0(x)$ is spherically symmetric, the integral $\int f(x) \rho_0(x) d^2x$ disappears. Therefore, the expectation value is the function evaluated at the center of the wavepacket:
\begin{equation*}
    \int_{\mathbb{R}^2} f(r) \rho_i(r) d^2r = f(r_i) \int_{\mathbb{R}^2} \rho_0(x) d^2x = f(r_i).
\end{equation*}
Substituting this result back into the distance formula gives:
\begin{align*}
    d_D(\omega_{r_1}, \omega_{r_2}) &\ge f(r_1) - f(r_2) \\
    &= \frac{r_1 - r_2}{\|r_1 - r_2\|} \cdot (r_1 - r_2) = \|r_1 - r_2\|.
\end{align*}
Thus, combining the lower and upper bound, we recover the classical Euclidean
distance \cite{D_Andrea_2010}:
\begin{equation*}
    d_D(\omega_{r_1}, \omega_{r_2}) = \|r_1 - r_2\|.
\end{equation*}

\subsection{Weyl Operators with the Free Dirac Operator} \label{weyl_free_proof}
To evaluate the Connes distance over the family of individual Weyl operators, we first calculate the expectation value of a Weyl operator $T(k,q) = e^{i(k\cdot \hat{r} + q\cdot\hat{p})} =
e^{ik\cdot\hat{r}}
e^{iq\cdot\hat{p}}
e^{\frac{i}{2}k\cdot q}$ 
on a localized state
\begin{equation*}
    \ket{\Psi_{r_1}} = \ket{\zeta_{r_1}} \otimes \ket{\chi}, \qquad \braket{\chi|\chi} = 1.
\end{equation*}
The momentum exponential acts as a spatial translation operator. Its action on the Gaussian wavepacket is $\braket{r | e^{iq\cdot\hat{p}} | \zeta_{r_1}} = \zeta_{r_1}(r + q)$. The expectation value $\omega_{r_1}(T) = \braket{\zeta_{r_1}|T(k,q)|\zeta_{r_1}}$ is the integral over position space:
\begin{align*}
    \omega_{r_1}(T) &= e^{\frac{i}{2}k\cdot q} \int_{\mathbb{R}^2} \zeta_{r_1}^*(r) e^{i k \cdot r} \zeta_{r_1}(r + q) \, d^2r \\
    &= \frac{e^{\frac{i}{2}k\cdot q}}{\pi \sigma^2} \int_{\mathbb{R}^2} e^{i k \cdot r} e^{-\frac{|r-r_1|^2}{2\sigma^2}} e^{-\frac{|r + q - r_1|^2}{2\sigma^2}} \, d^2r.
\end{align*}
Using the change of variables $x = r - r_1$ as we did previously gives:
\begin{align*}
    \omega_{r_1}(T) 
    &= \frac{e^{\frac{i}{2}k\cdot q} e^{i k \cdot r_1}}{\pi \sigma^2} \int_{\mathbb{R}^2} e^{i k \cdot x} e^{-\frac{|x|^2}{2\sigma^2}} e^{-\frac{|x + q|^2}{2\sigma^2}} \, d^2x \\
    &= \frac{e^{\frac{i}{2}k\cdot q} e^{i k \cdot r_1}}{\pi \sigma^2}
       \int_{\mathbb{R}^2} e^{i k \cdot x} e^{-\frac{|x|^2}{2\sigma^2}} \\
    &\qquad \times e^{-\frac{1}{2\sigma^2} \big( |x|^2 + 2x \cdot q + |q|^2 \big)} \, d^2x \\
    &= \frac{e^{\frac{i}{2}k\cdot q} e^{i k \cdot r_1}}{\pi \sigma^2} \int_{\mathbb{R}^2} e^{i k \cdot x} e^{-\frac{1}{2\sigma^2} \big( 2|x|^2 + 2x \cdot q + |q|^2 \big)} \, d^2x \\
    &= \frac{e^{\frac{i}{2}k\cdot q} e^{i k \cdot r_1}}{\pi \sigma^2} \int_{\mathbb{R}^2} e^{-\frac{1}{\sigma^2} |x|^2 + \left(i k - \frac{q}{\sigma^2}\right) \cdot x - \frac{|q|^2}{2\sigma^2}} \, d^2x \\
    &=  e^{i k \cdot r_1} e^{-\frac{\sigma^2 \|k\|^2}{4} - \frac{\|q\|^2}{4\sigma^2}}.
\end{align*}
The difference between two such states centered at $r_1$ and $r_2$ is:
\begin{equation*}
    |\omega_{r_1}(T) - \omega_{r_2}(T)| = e^{-\frac{\sigma^2 \|k\|^2}{4} - \frac{\|q\|^2}{4\sigma^2}} |e^{ik\cdot r_1} - e^{ik\cdot r_2}|.
\end{equation*}
For the free Dirac operator, the constraint is
\begin{equation*}
    \|[D_{\text{free}}, \pi(T)]\| \le 1.
\end{equation*}
It is equivalent to $\|k\| \le 1$, with no restriction on $q$.
\begin{equation*}
    d_{D_\text{free}}(\zeta_1,\zeta_2) = \sup_{k,q} \frac{1}{\|k\|} |e^{ik\cdot r_1} - e^{ik\cdot r_2}| \,  e^{-\frac{\sigma^2 \|k\|^2}{4} - \frac{\|q\|^2}{4\sigma^2}}.
\end{equation*}
To maximize the distance, the supremum selects $q = 0$ to eliminate the exponential decay. It is also maximized when $k \to 0$. We decompose it as a constant times a unit vector $k = \alpha \Vec{n}$ as $\alpha \to 0$:
\begin{equation*}
    \lim_{\alpha \to 0} \frac{1}{\alpha} | e^{i\alpha \Vec{n}\cdot r_1} - e^{i\alpha\Vec{n}\cdot r_2} | = | i \Vec{n} \cdot (r_1-r_2) |.
\end{equation*}
To maximize this dot product, we choose $\Vec{n} = \frac{r_1 - r_2}{\|r_1 - r_2\|}$ which gives $\|r_1 - r_2\|$

Bounding the numerator gives the upper bound:
\begin{equation*}
    \frac{|e^{ik\cdot r_1}-e^{ik\cdot r_2}|}{\|k\|} \le \frac{|k\cdot(r_1-r_2)|}{\|k\|} \le \|r_1-r_2\|.
\end{equation*}
Combining the lower bound and the upper bound, we recover the equality $d_{D_\text{free}}(\zeta_1,\zeta_2) = \|r_1 - r_2\|_2$.

\subsection{Weyl Operators with the Magnetic Dirac Operator} \label{weyl-magnetic-dirac-proof}

We evaluate the commutator of the magnetic Dirac operator $D_B$ with the Weyl operator $T(k,q)$.
It decomposes linearly into momentum and gauge potential terms:
\begin{equation*}
    [\Pi_j, T(k,q)] = [p_j, T(k,q)] - [A_j(\hat{r}), T(k,q)].
\end{equation*}
Using $[p_j, e^{ik_j\hat{r}_j}] = k_j e^{ik_j\hat{r}_j}$, the momentum term simplifies to $[p_j, T(k,q)] = k_j T(k,q)$.

To evaluate the gauge potential commutator, we use the following relation:
\begin{equation*}
    e^{iq\cdot\hat{p}} A_j(\hat{r}) e^{-iq\cdot\hat{p}} = A_j(\hat{r} + q).
\end{equation*}
Applying this to the Weyl operator
\begin{equation*}
    T(k,q) = e^{ik\cdot\hat{r}} e^{iq\cdot\hat{p}} e^{\frac{i}{2}k\cdot q},
\end{equation*}
we can commute the gauge field through $T(k,q)$:
\begin{align*}
    T(k,q) A_j(\hat{r}) &= e^{ik\cdot\hat{r}} e^{iq\cdot\hat{p}} A_j(\hat{r}) e^{\frac{i}{2}k\cdot q} \\
    &= e^{ik\cdot\hat{r}} A_j(\hat{r} + q) e^{iq\cdot\hat{p}} e^{\frac{i}{2}k\cdot q} \\
    &= A_j(\hat{r} + q) T(k,q).
\end{align*}
Therefore, the commutator with the gauge field is:
\begin{align*}
    -[A_j(\hat{r}), T(k,q)] &= -A_j(\hat{r})T(k,q) + T(k,q)A_j(\hat{r}) \\
    &= \big( -A_j(\hat{r}) + A_j(\hat{r} + q) \big) T(k,q).
\end{align*}
We substitute the Landau gauge $\vec{A} = (-By, 0)$:
\begin{align*}
    -A_x(\hat{y}) + A_x(\hat{y} + q_y) &= B\hat{y} - B(\hat{y} + q_y) = -Bq_y, \\
    -A_y(\hat{x}) + A_y(\hat{x} + q_x) &= 0.
\end{align*}
Therefore, in that gauge, we have:
\begin{align*}
    [\Pi_x, T(k,q)] &= (k_x - Bq_y) T(k,q), \\
    [\Pi_y, T(k,q)] &= k_y T(k,q).
\end{align*}
Since the Weyl operators act as the identity on the coin space, the mass term $m\sigma_y$ commutes with $\pi(T)$. Thus, the full commutator:
\begin{equation*}
    [D_B, \pi(T)] = \big( \sigma_z(k_x - Bq_y) + \sigma_x k_y \big) T(k,q).
\end{equation*}
The commutator's norm is then:
\begin{equation*}
    \| [D_B, \pi(T)] \| = \sqrt{(k_x - Bq_y)^2 + k_y^2}.
\end{equation*}

\section{Quantum Walk Action on the Coin}
Here, we prove Eq. \ref{qw-action-coin} by applying sequentially the matrices of the walk operator to a state $\ket{r} \otimes \ket{c}$:
\begin{enumerate}
\item First shift ($T_x$): The walker shifts to $r + s_x e_x$, conditioned by $P_{s_x}$.
\item First gauge phase ($D_x$): The walker picks up the phase at its new position, applying the diagonal unitary $e^{i \epsilon A_x(r + s_x e_x) \sigma_z}$.
\item First coin mixing ($C_1$): The coin state is mixed by applying $H$.
\item Second shift ($T_y$): The walker shifts along the $y$ axis to the final position $r + s_x e_x + s_y e_y = r + \delta$, conditioned by $P_{s_y}$.
\item Second gauge phase ($D_y$): The walker picks up the phase at its final position, applying $e^{i \epsilon A_y(r + \delta) \sigma_z}$.
\item Second coin mixing ($C_2$): $M_\epsilon H$ is applied.
\end{enumerate}
We obtain 
\[
U_\delta(r) = M_\epsilon H \, e^{i \epsilon A_y(r + \delta) \sigma_z} P_{s_y} \, H \, e^{i \epsilon A_x(r + s_x e_x) \sigma_z} P_{s_x}.
\]

\section{The Quantum Walk's Spread}
\subsection{Thin Ring Density Approximation} \label{bessel-proof}
The density of the thin ring is approximated in polar coordinates $(r, \theta)$ by $\rho_t(r, \theta) = \frac{1}{2\pi R} \delta(r - R)$. Its integral is:
\begin{equation*}
    \int e^{i k \cdot r} \rho_t(r) d^2r.
\end{equation*}
In polar coordinates, the momentum and position vectors are
\begin{align*}
    k &= (\|k\|\cos\phi, \|k\|\sin\phi), \\
    r &= (\|r\|\cos\theta, \|r\|\sin\theta).
\end{align*}
The dot product is $k \cdot r = \|k\|\|r\|\cos(\theta - \phi)$, and the spatial area element is $d^2r = r dr d\theta$.

The integral for the point-source expansion becomes: 
\[
\int_0^{2\pi} \int_0^\infty e^{i \|k\| \|r\| \cos(\theta - \phi)} \left( \frac{1}{2\pi R} \delta(r - R) \right) r dr d\theta.
\]
Integrating over $r$ forces $r = R$ everywhere inside the expression
\[
\frac{1}{2\pi} \int_0^{2\pi} e^{i ||k|| R \cos(\theta - \phi)} d\theta = \frac{1}{2\pi} \int_0^{2\pi} e^{i \|k\| R \cos\theta} d\theta .
\]
By periodicity and angular symmetry, the phase shift $\phi$ can be eliminated. Finally, the integral is equal to the Bessel function of the first kind of order zero:
\[
\frac{1}{2\pi} \int_0^{2\pi} e^{i \|k\| R \cos\theta} d\theta  = J_0(\|k\|R).
\]

\subsection{1+1D Plastic Quantum Walk} \label{plastic-qw}
For the 1+1D plastic walk $W_\epsilon = \Lambda^{-1} S C_{-\zeta} S C_\zeta \Lambda$, the sequential application of the spin-dependent shifts $S$ expands the spatial support of a localized state by $\delta \in \{-2, 0, 2\}$. Applying the walk to a basis state $\ket{x, c}$ gives:
\begin{equation*}
    W_\epsilon \ket{x, c} = \ket{x-2} \otimes T_{-2} \ket{c} + \ket{x} \otimes T_0 \ket{c} + \ket{x+2} \otimes T_2 \ket{c},
\end{equation*}
where the coin operators for the left and right maximal jumps are given by:
\begin{align*} 
T_{-2} &= \Lambda^{-1} P_{-1} C_{-\zeta} P_{-1} C_\zeta \Lambda, \\ 
T_2 &= \Lambda^{-1} P_1 C_{-\zeta} P_1 C_\zeta \Lambda.
\end{align*}
with $P_{-1}, P_1$ the standard orthogonal projectors onto the coin basis.

For a spatial test observable $a$ represented by $\pi(a) = \sum_x a_x \ket{x}\bra{x} \otimes \mathbb{I}_2$, we recall the discrete gradient $\Delta_x = a_{x+2} - a_x$. The action of the commutator on a localized state is: \begin{equation*}
    [W_\epsilon, \pi(a)] \ket{x, c} = \Delta_{x-2} \ket{x-2} T_{-2} \ket{c} - \Delta_x \ket{x+2} T_2 \ket{c}.
\end{equation*}
Since $T_{-2}$ and $T_2$ project into orthogonal subspaces, the cross-terms in the squared norm vanish. Using the fact that $\| P_{-1} C_{-\zeta} P_{-1} \| = \| P_1 C_{-\zeta} P_1 \| = |\cos \theta|$, the upper bound of the global operator norm is restricted by the maximum spatial gradient:
\begin{equation*}
    \| [W_\epsilon, \pi(a)] \| \le |\cos \theta| \max_x |\Delta_x|.
\end{equation*}
\textit{Proof}:
\begin{align*}
    &\|[W_\epsilon,\pi(a)]\ket{\psi}\|^2 \\
    &= \sum_k \|\Delta_k T_{-2} \ket{c_{k+2}} \|^2 + \| \Delta_{k-2} T_2 \ket{c_{k-2}} \|^2 \\
    &= \sum_x \|\Delta_x T_{-2} \ket{c_{x+2}} \|^2 + \| \Delta_{x} T_2 \ket{c_{x}} \|^2 \\
    &= \sum_x |\Delta_x|^2 \big(\|T_{-2}\ket{c_{x+2}} \|^2 + \|T_2 \ket{c_{x}}\|^2 \big) \\
    &= \sum_x |\Delta_x|^2 \big(\|\Lambda^{-1} P_{-1} C_{-\zeta} P_{-1} C_\zeta \Lambda \ket{c_{x+2}} \|^2 \\
    &\qquad + \|\Lambda^{-1} P_1 C_{-\zeta} P_1 C_\zeta \Lambda \ket{c_{x}}\|^2 \big) \\
    &= \sum_x |\Delta_x|^2 \big(\| P_{-1} C_{-\zeta} P_{-1} \ket{c'_{x+2}} \|^2 \\
    &\qquad + \| P_1 C_{-\zeta} P_1 \ket{c'_{x}}\|^2 \big) \\
    &= \sum_x |\Delta_x|^2 |\cos\theta|^2 \big(\|P_{-1}\ket{c'_{x+2}}\|^2 + \|P_1\ket{c'_x}\|^2 \big) \\
    &\leq \max_x |\Delta_x|^2 |\cos\theta|^2 \\
    &\qquad \times \sum_x \big(\|P_{-1}\ket{c'_{x+2}}\|^2 + \|P_1\ket{c'_x}\|^2 \big) \\
    &= \max_x |\Delta_x|^2 |\cos\theta|^2 \\
    &\qquad \times \sum_x \big(\|P_{-1}\ket{c'_{x}}\|^2 + \|P_1\ket{c'_x}\|^2 \big) \\
    &= \max_x |\Delta_x|^2 |\cos\theta|^2 \sum_x \|\ket{c'_x}\|^2 \\
    &= \max_x |\Delta_x|^2 |\cos\theta|^2.
\end{align*}
To prove that this upper bound is in fact an exact equality, we show that there exists a normalized state in the Hilbert space that saturates it. 
The definition of the operator norm guarantees that $\| [W_\epsilon, \pi(a)] \| \ge \| [W_\epsilon, \pi(a)] \ket{\psi^\star} \|$ for any state $\ket{\psi^\star}$ with $\braket{\psi^\star|\psi^\star} = 1$.
Let $x^\star$ be the specific lattice site that maximizes the spatial gradient, such that $|\Delta_{x^\star}| = \max_x |\Delta_x|$. 
We build a trial state localized at this site: $\ket{\psi^\star} = \ket{x^\star} \otimes \ket{c^\star}$, where $\ket{c^\star}$ is a normalized coin state to be determined.
Applying the commutator to this state and squaring it gives:
\begin{equation*}
    \begin{split}
    \| [W_\epsilon, \pi(a)] \ket{\psi^\star} \|^2
      &= |\Delta_{x^\star-2}|^2 \| T_{-2} \ket{c^\star} \|^2 \\
      &\quad + |\Delta_{x^\star}|^2 \| T_2 \ket{c^\star} \|^2.
    \end{split}
\end{equation*}
To isolate $\Delta_{x^\star}$ and eliminate the $\Delta_{x^\star-2}$ term, we require $T_{-2} \ket{c^\star} = 0$. Recall the coin operators:
\begin{align*} 
T_{-2} &= \Lambda^{-1} P_{-1} C_{-\zeta} P_{-1} (C_\zeta \Lambda), \\
T_2 &= \Lambda^{-1} P_1 C_{-\zeta} P_1 (C_\zeta \Lambda).
\end{align*}
The rightmost pair of operators $C_\zeta \Lambda$ forms a unitary matrix. We can therefore define our initial coin state such that this unitary rotates it into the $P_1$ subspace (which corresponds to the coin state $\ket{0}$). Let
\begin{equation*}
    \ket{c^\star} = \Lambda^\dagger C_\zeta^\dagger \ket{0}.
\end{equation*}
Substituting this back, we first evaluate the rightmost action: $C_\zeta \Lambda \ket{c^\star} = \ket{0}$. The projector $P_{-1}$ acts on this state, giving $P_{-1} \ket{0} = 0$, which leads to $T_{-2} \ket{c^\star} = 0$. 
For the right-moving branch, the projector leaves the state invariant $P_1 \ket{0} = \ket{0}$. The remaining operation gives:
\begin{equation*}
    \| T_2 \ket{c^\star} \| = \| \Lambda^{-1} P_1 C_{-\zeta} P_1 \ket{0} \| = |\cos \theta|.
\end{equation*}

Substituting this back into the squared norm expression, the norm evaluated on our trial state is: 
\begin{equation*}
    \| [W_\epsilon, \pi(a)] \ket{\psi^\star} \| = |\Delta_{x^\star}| |\cos \theta| = \max_x |\Delta_x| |\cos \theta|.
\end{equation*}
The global operator norm cannot be less than the norm evaluated on any valid normalized state, this gives the lower bound: $\| [W_\epsilon, \pi(a)] \| \ge |\cos \theta| \max_x |\Delta_x|$. Coupled with the upper bound obtained previously, this gives the equality:
\begin{equation*}
    \| [W_\epsilon, \pi(a)] \| = |\cos \theta|\max_x |\Delta_x|.
\end{equation*}

Therefore, the constraint $\| [W_\epsilon, \pi(a)] \| \le 2\epsilon$ is equivalent to $\max_x |\Delta_x| \le \frac{2\epsilon}{|\cos \theta|}$.

\textit{Remark: The walk connects only sites of the same parity, thus the lattice is split into two independent sublattices, even and odd. The initial state at $x=0$ evolves within the even sublattice, and all distances considered here are therefore restricted to that sublattice.}

\section{Numerical Methods}
We use two numerical approaches to evaluate the distance. In $2+1$ dimensions, we search over a finite set of individual Weyl operators and estimate their commutator norms. In $1+1$ dimensions, we optimize over all real diagonal observables on the occupied parity sector of a finite lattice. The first approach is a restricted search, the second leads to a convex optimization problem that is solved to finite numerical precision.

\subsection{2+1D Quantum Walk} \label{numerical-details}

\subsubsection{Lattice and Search Domain}
The state is represented as a complex array of shape $(L,L,2)$ on a periodic square lattice. We use $L=64$, $\epsilon=0.01$, $m=1$, and $N=1,\ldots,20$. The initial Gaussian has width $\sigma_{\mathrm{lat}}=\sqrt{L/(4\pi)}\simeq2.26$ lattice sites, corresponding to the physical width $\sigma=\epsilon\sigma_{\mathrm{lat}}$. The same initial envelope is used for the free and magnetic evolutions. For the magnetic case, the plaquette flux is $\Phi=B_{\mathrm{phys}}\epsilon^2=2\pi/L$.

In this subsection, $n$ and $q$ denote integer lattice coordinates and shifts, and $k$ denotes the dimensionless lattice momentum. Thus $r=\epsilon n$, $q_{\mathrm{phys}}=\epsilon q$, and $k=\epsilon k_{\mathrm{phys}}$. We use the unscaled test operator
\begin{equation*}
    A_{k,q}=V_kS_q\otimes\mathbb I_2,
    \qquad (A_{k,q}\psi)(n)=e^{ik\cdot n}\psi(n-q).
\end{equation*}
All array shifts are periodic. The sampled momenta are
\begin{equation*}
    k=\frac{2\pi}{L}(j_x,j_y),
    \qquad j_x,j_y\in\{-16,\ldots,16\}.
\end{equation*}
These values are compatible with the lattice periodicity, but they cover only part of the reciprocal lattice. In the free calculation we take $q=(0,0)$, which is consistent with the position-observable restriction discussed in Sec.~\ref{discrete-comm-algebra}. At step $N$, the magnetic calculation searches $q=(0,q_y)$ with $q_y=-N,\ldots,N$.

\subsubsection{Expectation Differences and Commutator Norms}
Let $\psi_N=W^N\psi_0$, where $W=W_\epsilon$ or $W_\epsilon^B$ for the chosen run. We define
\begin{align*}
    b_N(k,q)&=\left|\braket{\psi_N|A_{k,q}|\psi_N}
                  -\braket{\psi_0|A_{k,q}|\psi_0}\right|,\\
    C_{k,q}&=WA_{k,q}-A_{k,q}W.
\end{align*}
When $\|C_{k,q}\|>0$, we scale the observable as $\pi(a)=cA_{k,q}$ and choose $c$ to saturate the constraint $c\|C_{k,q}\| \leq \epsilon$. This gives the candidate distance
\begin{equation} \label{app-weyl-candidate}
    d_N(k,q)=\epsilon\frac{b_N(k,q)}{\|C_{k,q}\|}.
\end{equation}
For each translation, the code first builds the spatial overlap array
\begin{equation*}
    \begin{split}
    h_{N,q}(n)=\sum_{s=0}^1\big[&\psi_{N,s}^*(n)\psi_{N,s}(n-q)\\
                             &-\psi_{0,s}^*(n)\psi_{0,s}(n-q)\big].
    \end{split}
\end{equation*}
The numerator is then evaluated for each sampled momentum according to
\begin{equation*}
    b_N(k,q)=\left|\sum_n e^{ik\cdot n}h_{N,q}(n)\right|.
\end{equation*}
Thus, the spatial part is computed once for each translation, while the phase sum retains the dependence on $k$.

The walk, the test operators, and their adjoints are applied using local coin operations, phase arrays, and periodic shifts. We therefore avoid constructing the full $2L^2\times2L^2$ matrices explicitly. Before computing the evolution, we test the implemented adjoints numerically through
\begin{equation*}
    \braket{u|\mathcal O v}=\braket{\mathcal O^\dagger u|v}
\end{equation*}
on normalized random vectors for both the walk and the sampled test operators. Any discrepancy larger than $10^{-11}$ is rejected. This is a numerical check on the implemented adjoints for the sampled vectors, rather than an exhaustive test of the corresponding operator identities.

\subsubsection{Candidate Ranking and Numerical Refinement}
A direct evaluation of the operator norm for every $(k,q)$ pair would be relatively expensive. We therefore use a cheaper estimate to rank the candidates. The code applies $C_{k,q}$ to 16 normalized states: nine arrays with plane-wave phases and seven complex Gaussian random vectors. The random vectors are generated with seed 123. We define
\begin{equation*}
    \ell(k,q)=\max_j\|C_{k,q}u_j\|\leq\|C_{k,q}\|.
\end{equation*}
Thus, when $\ell(k,q)>0$, the quantity $\epsilon b_N(k,q)/\ell(k,q)$ provides an upper bound on the corresponding candidate distance. To avoid numerical problems when the denominator becomes very small, we impose a lower cutoff of $10^{-12}$ on the denominator and discard numerators below the same threshold.

For each searched translation, candidates are ranked by this ratio and the top 20 are refined using
\begin{equation*}
    \|C_{k,q}\|=\sqrt{\lambda_{\max}(C_{k,q}^\dagger C_{k,q})}.
\end{equation*}
The positive operator is passed to \texttt{scipy.sparse.linalg.eigsh} as a \texttt{LinearOperator}. We request its largest eigenvalue with tolerance $10^{-8}$ and allow at most 250 iterations. The eigensolver is initialized with a normalized complex random vector generated using seed 42. Since $W$ and $A_{k,q}$ are independent of the evolution step, the norm estimates are cached and reused. The reported distance is the largest candidate ratio obtained after this refinement, up to the numerical error of the eigensolver.

Evaluating only the top 20 candidates does not by itself certify the global maximum, since some unrefined candidates could in principle be larger. We therefore check the stability of the result separately by varying the momentum grids, candidate pools, and lattice sizes.

\subsubsection{Kernel Observables and Dynamical Consistency}
If $[W,A]=0$, unitary evolution leaves the expectation value of $A$ unchanged, so its numerator vanishes. More generally, telescoping gives
\begin{align*}
    W^{-N}AW^N-A
      &=\sum_{j=0}^{N-1}W^{-j}(W^{-1}AW-A)W^j,\\
    \left|\omega_N(A)-\omega_0(A)\right|
      &\leq N\|[W,A]\|.
\end{align*}
It follows that every admissible candidate in Eq.~\ref{app-weyl-candidate} satisfies $d_N(k,q)\leq N\epsilon$. Exact kernel operators contribute zero to the defining supremum, so their resulting $0/0$ ratio is simply omitted.

\subsection{1+1D Plastic Quantum Walk} \label{app-1+1-qw}

\subsubsection{Evolution and Parity Reduction}
The massless walk is implemented as
\begin{equation*}
    W=\Lambda^{-1}SC_{-\zeta}SC_\zeta\Lambda,
    \qquad \zeta=0,
\end{equation*}
using the coin and $\Lambda$ matrices defined in the main text, with $c(x)=\cos\theta(x)$. The shift convention is
\begin{equation*}
    S\ket{x,0}=\ket{x+1,0},\qquad
    S\ket{x,1}=\ket{x-1,1},
\end{equation*}
where $x$ is the integer site coordinate and the corresponding physical position is $\epsilon x$. In the numerical implementation, the right and left shifts are performed with positive and negative array rolls. The initial state is localized at $x=0$ with coin state $(\ket{0}+i\ket{1})/\sqrt{2}$.

On an even periodic lattice of $L$ sites, one complete step preserves position parity. A walker initially at the origin therefore visits only half of the lattice sites. We restrict the optimization to these $L/2$ active sites. Including the two coin components gives a reduced Hilbert space of dimension $L$. The sparse walk matrix is built by applying the same evolution routine to each basis vector in the active sector and discarding entries with magnitude below $10^{-13}$. We check that this reduced matrix remains unitary within standard numerical tolerance.

\subsubsection{Finite-Lattice Convex Problem}
Let $a\in\mathbb R^{L/2}$ denote the observable values on the active sites, and
\begin{equation*}
    A(a)=\operatorname{diag}(a)\otimes\mathbb I_2,
    \qquad \Delta P_N(x)=P_N(x)-P_0(x).
\end{equation*}
The finite-lattice problem is
\begin{equation} \label{app-plastic-convex}
    \begin{aligned}
    \text{maximize}\quad &\sum_{x\ \mathrm{active}}\Delta P_N(x)a_x\\
    \text{subject to}\quad &\|WA(a)-A(a)W\|_2\leq2\epsilon,\\
                         &a_0=0.
    \end{aligned}
\end{equation}
The feasible set is invariant under $a\mapsto-a$, so maximizing the expectation difference is equivalent to taking the supremum of its absolute value. Also, adding a constant to $a$ leaves both the commutator and the expectation difference unchanged for normalized states. We therefore fix this freedom by imposing $a_0=0$, this acts purely as a mathematical reference point. In the numerical calculation, we also subtract the mean of $\Delta P_N$ to remove the small normalization drift introduced by floating-point arithmetic.

The objective function is linear and the constraint is convex. We solve this finite-dimensional problem with \texttt{CVXPY}, using the \texttt{CLARABEL} solver when available and \texttt{SCS} otherwise. The solver is called with its default tolerances and with a warm start. The observable obtained at the previous step is retained as the initial value, although whether this is used internally depends on the solver interface.

\subsubsection{Accuracy Checks and Reported Calculations}
The current routine accepts both \texttt{OPTIMAL} and \texttt{OPTIMAL\_INACCURATE} statuses and stops with an error for any other status. We independently evaluate
\begin{equation*}
    \nu_N=\|WA(a_N)-A(a_N)W\|_2
\end{equation*}
alongside normalization, parity leakage, and the condition $a_0$. These checks verify that the returned observable satisfies the imposed constraint to the tested tolerance, but they do not by themselves establish global optimality.

For the homogeneous plots, we evolve the probability distribution directly and evaluate the distance from the derived expression
\begin{equation*}
    d_N=\frac{\epsilon}{|\cos\theta|}\sum_x|x|P_N(x).
\end{equation*}
These curves are therefore not obtained from a separate convex optimization at each time step. We use $L=800$, $N\leq12$, $\epsilon=1$, and $\theta\in{\pi/16,\pi/4,3\pi/8}$.
On a periodic lattice, the infinite-line expression requires that the relevant positions lie within the region where distance from the origin agrees with $|x|$.

For the inhomogeneous case, Eq.~\ref{app-plastic-convex} is solved at every step. The parameters are $L=200$, $N\leq30$, and $\epsilon=1$. We consider the four profiles
\begin{equation*}
\begin{aligned}
    c_{\mathrm{hom}}(x)&=0.70,\\
    c_{\mathrm{slow}}(x)&=0.85-0.50e^{-x^2/(2\cdot9^2)},\\
    c_{\mathrm{fast}}(x)&=0.30+0.50e^{-x^2/(2\cdot9^2)},\\
    c_{\mathrm{int}}(x)&=0.60-0.25\tanh(x/5).
\end{aligned}
\end{equation*}
We also perform a separate check of the corrected shift implementation using $L=100$, $N=8$, $\epsilon=1$, and $\theta=\pi/4$. With \texttt{CVXPY} 1.9.2 and \texttt{CLARABEL}, the optimized distance is approximately $11.3137084826$, compared with $11.3137084990$ from the homogeneous expression. The relative difference is $1.45\times10^{-9}$. The solver reports \texttt{OPTIMAL}, and the independently evaluated commutator norm is approximately $1.9999999984$, below the constraint limit of 2.

Finally, periodic boundary conditions affect both the wavepacket dynamics and the global observable constraint. For this reason, checking that the wavepacket itself remains localized is not sufficient to prove convergence to the infinite-lattice result.

\bibliography{apssamp}

@PREAMBLE{
 "\providecommand{\noopsort}[1]{}" 
 # "\providecommand{\singleletter}[1]{#1}%" 
}

@article{Connes1995,
  author    = {Connes, Alain},
  title     = {Noncommutative geometry and reality},
  journal   = {Journal of Mathematical Physics},
  volume    = {36},
  number    = {11},
  pages     = {6194--6231},
  year      = {1995},
  publisher = {AIP Publishing},
  doi       = {10.1063/1.531241}
}

@article{Cagnache_2011,
  title     = {The spectral distance in the {M}oyal plane},
  volume    = {61},
  number    = {10},
  issn      = {0393-0440},
  doi       = {10.1016/j.geomphys.2011.04.021},
  journal   = {Journal of Geometry and Physics},
  publisher = {Elsevier BV},
  author    = {Cagnache, Eric and D'Andrea, Francesco and Martinetti, Pierre and Wallet, Jean-Christophe},
  year      = {2011},
  month     = oct,
  pages     = {1881--1897}
}

@article{2020,
  title     = {The {N}oncommutative {G}eometry of the {L}andau {H}amiltonian: {M}etric {A}spects},
  issn      = {1815-0659},
  doi       = {10.3842/sigma.2020.146},
  journal   = {Symmetry, Integrability and Geometry: Methods and Applications},
  publisher = {SIGMA},
  author    = {De Nittis, Giuseppe and Sandoval, Maximiliano},
  year      = {2020},
  month     = dec
}

@misc{dimolfetta2019quantumwalkcontinuoustimecontinuousspacetime,
  title         = {A quantum walk with both a continuous-time and a continuous-spacetime limit}, 
  author        = {Giuseppe Di Molfetta and Pablo Arrighi},
  year          = {2019},
  eprint        = {1906.04483},
  archivePrefix = {arXiv},
  primaryClass  = {quant-ph},
  url           = {https://arxiv.org/abs/1906.04483}
}

@incollection{konno2008quantum,
  author    = {Konno, Norio},
  title     = {Quantum {W}alks},
  booktitle = {Quantum Potential Theory},
  series    = {Lecture Notes in Mathematics},
  volume    = {1954},
  pages     = {309--452},
  year      = {2008},
  publisher = {Springer},
  address   = {Berlin, Heidelberg},
  doi       = {10.1007/978-3-540-79992-4_3},
  url       = {https://springer.com}
}

@misc{konno2005newtypelimittheorems,
  title         = {A {N}ew {T}ype of {L}imit {T}heorems for the {One-Dimensional} {Quantum Random Walk}}, 
  author        = {Norio Konno},
  year          = {2005},
  eprint        = {quant-ph/0206103},
  archivePrefix = {arXiv},
  primaryClass  = {quant-ph},
  url           = {https://arxiv.org/abs/quant-ph/0206103}
}

@article{PhysRevA.88.042301,
  title     = {Quantum walks as massless {D}irac fermions in curved space-time},
  author    = {Di Molfetta, Giuseppe and Brachet, M. and Debbasch, Fabrice},
  journal   = {Phys. Rev. A},
  volume    = {88},
  number    = {4},
  pages     = {042301},
  year      = {2013},
  month     = oct,
  publisher = {American Physical Society},
  doi       = {10.1103/PhysRevA.88.042301},
  url       = {https://link.aps.org/doi/10.1103/PhysRevA.88.042301}
}

@article{Arrighi2018quantumwalkingin,
  title     = {Quantum walking in curved spacetime: discrete metric},
  author    = {Arrighi, Pablo and Molfetta, Giuseppe Di and Facchini, Stefano},
  journal   = {{Quantum}},
  issn      = {2521-327X},
  publisher = {{Verein zur F{\"{o}}rderung des Open Access Publizierens in den Quantenwissenschaften}},
  volume    = {2},
  pages     = {84},
  year      = {2018},
  month     = aug,
  doi       = {10.22331/q-2018-08-22-84},
  url       = {https://doi.org/10.22331/q-2018-08-22-84}
}

@article{Arrighi_Nesme_Forets_2014,
  author      = {Arrighi, Pablo and Forets, Marcelo and Nesme, Vincent},
  title       = {{The Dirac Equation as a Quantum Walk: Higher Dimensions, Observational Convergence}},
  journal     = {Journal of Physics A: Mathematical and Theoretical},
  volume      = {47},
  number      = {46},
  pages       = {465302},
  year        = {2014},
  doi         = {10.1088/1751-8113/47/46/465302},
  eprint      = {1307.3524},
  eprinttype  = {arXiv},
  eprintclass = {quant-ph}
}

@article{PhysRevA.97.062111,
  author  = {Arrighi, Pablo and Di Molfetta, Giuseppe and M\'arquez-Mart\'{\i}n, Iv\'an and P\'erez, Armando},
  title   = {{Dirac Equation as a Quantum Walk over the Honeycomb and Triangular Lattices}},
  journal = {Physical Review A},
  volume  = {97},
  number  = {6},
  pages   = {062111},
  year    = {2018},
  month   = jun,
  doi     = {10.1103/PhysRevA.97.062111}
}

@phdthesis{Arnault_2021,
  author      = {Arnault, Pablo},
  title       = {Discrete-Time Quantum Walks and Gauge Theories},
  institution = {Universit\'e Pierre et Marie Curie},
  address     = {Paris, France},
  year        = {2017},
  eprint      = {1710.11123},
  eprinttype  = {arXiv},
  eprintclass = {quant-ph},
  url         = {https://arxiv.org/abs/1710.11123}
}

@book{Connes_1994,
  author    = {Connes, Alain},
  title     = {Noncommutative Geometry},
  publisher = {Academic Press},
  address   = {San Diego},
  year      = {1994},
  isbn      = {978-0-12-185860-5}
}

@book{Suijlekom_2023a,
  author    = {van Suijlekom, Walter D.},
  title     = {Noncommutative Geometry and Particle Physics},
  edition   = {2},
  series    = {Mathematical Physics Studies},
  publisher = {Springer},
  address   = {Cham},
  year      = {2025},
  doi       = {10.1007/978-3-031-59120-4}
}

@inproceedings{hijazi,
  author    = {Hijazi, Ousama},
  title     = {Spectral properties of the {D}irac operator and geometrical structures},
  booktitle = {Geometric Methods for Quantum Field Theory},
  pages     = {76--115},
  year      = {2001},
  publisher = {World Scientific}
}

@article{Zak1964MAGNETICTG,
  title     = {Magnetic {T}ranslation {G}roup},
  author    = {Zak, J.},
  journal   = {Phys. Rev.},
  volume    = {134},
  number    = {6A},
  pages     = {A1602--A1606},
  year      = {1964},
  month     = jun,
  publisher = {American Physical Society},
  doi       = {10.1103/PhysRev.134.A1602},
  url       = {https://link.aps.org/doi/10.1103/PhysRev.134.A1602}
}

@article{D_Andrea_2010,
  author  = {D'Andrea, Francesco and Martinetti, Pierre},
  title   = {A View on Optimal Transport from Noncommutative Geometry},
  journal = {SIGMA},
  volume  = {6},
  pages   = {057},
  year    = {2010},
  doi     = {10.3842/SIGMA.2010.057},
  eprint  = {0906.1267},
  archivePrefix = {arXiv},
  primaryClass  = {math.OA}
}

@Book{varilly2006introductionnoncommutativegeometry,
  author        = {Joseph C. Várilly},
  publisher     = {European Mathematical Society},
  title         = {{An Introduction to Noncommutative Geometry}},
  year          = {2006},
  archiveprefix = {arXiv},
  eprint        = {physics/9709045v1},
  primaryclass  = {math-ph},
}

@InProceedings{Reyes-Lega2016,
  author    = {Reyes-Lega, A. F.},
  booktitle = {Geometric, Algebraic and Topological Methods for Quantum Field Theory},
  title     = {Some {A}spects of {O}perator {A}lgebras in {Q}uantum {P}hysics},
  year      = {2016},
  editor    = {Cano, L. and Cardona, A. and Ocampo, H. and Reyes-Lega, A.},
  pages     = {1--74},
  publisher = {World Scientific},
  doi       = {10.1142/9789814730884_0001},
  url       = {http://www.worldscientific.com/doi/suppl/10.1142/9861/suppl_file/9861_chap01.pdf},
}

@Book{GraciaBondia2001,
  author    = {Gracia-Bond\'{i}a, J. M. and V\'{a}rilly, J. C. and Figueroa, H.},
  publisher = {Birkh\"auser},
  title     = {Elements of {N}oncommutative {G}eometry},
  year      = {2001},
  doi       = {10.1007/978-1-4612-0005-5},
}

@article{De_Palma_2021,
   title={The {Q}uantum {W}asserstein {D}istance of {O}rder 1},
   volume={67},
   ISSN={1557-9654},
   url={http://dx.doi.org/10.1109/TIT.2021.3076442},
   DOI={10.1109/tit.2021.3076442},
   number={10},
   journal={IEEE Transactions on Information Theory},
   publisher={Institute of Electrical and Electronics Engineers (IEEE)},
   author={De Palma, Giacomo and Marvian, Milad and Trevisan, Dario and Lloyd, Seth},
   year={2021},
   month=Oct, pages = {6627--6643} }

@article{Trevisan2025,
  author  = {Trevisan, Dario},
  title   = {Quantum optimal transport: an invitation},
  journal = {Bollettino dell'Unione Matematica Italiana},
  volume  = {18},
  pages   = {347--360},
  year    = {2025},
  doi     = {10.1007/s40574-024-00428-5}
}

@article{Bravyi_2006,
   title={{Lieb-Robinson Bounds and the Generation of Correlations and Topological Quantum Order}},
   volume={97},
   ISSN={1079-7114},
   url={http://dx.doi.org/10.1103/PhysRevLett.97.050401},
   DOI={10.1103/physrevlett.97.050401},
   number={5},
   journal={Physical Review Letters},
   publisher={American Physical Society (APS)},
   author={Bravyi, S. and Hastings, M. B. and Verstraete, F.},
   year={2006} }

@article{DimakisMuellerHoissen1998,
  author  = {Dimakis, Aristophanes and M{\"u}ller-Hoissen, Folkert},
  title   = {Connes' distance function on one-dimensional lattices},
  journal = {International Journal of Theoretical Physics},
  volume  = {37},
  number  = {3},
  pages   = {907--913},
  year    = {1998},
  doi     = {10.1023/A:1026684917059},
  eprint  = {q-alg/9707016},
  archivePrefix = {arXiv},
  primaryClass = {q-alg}
}

@misc{Besnard2021,
  author        = {Besnard, Fabien},
  title         = {Estimating noncommutative distances on graphs},
  year          = {2021},
  eprint        = {2105.09056},
  archivePrefix = {arXiv},
  primaryClass  = {math.OA},
  url           = {https://arxiv.org/abs/2105.09056}
}

@article{MartinettiTomassini2013,
  author  = {Martinetti, Pierre and Tomassini, Luca},
  title   = {Noncommutative Geometry of the Moyal Plane: Translation Isometries, Connes' Distance on Coherent States, Pythagoras Equality},
  journal = {Communications in Mathematical Physics},
  volume  = {323},
  number  = {1},
  pages   = {107--141},
  year    = {2013},
  doi     = {10.1007/s00220-013-1760-8},
  eprint  = {1110.6164},
  archivePrefix = {arXiv},
  primaryClass = {math-ph}
}

\end{document}